# FIRST: A Framework for Optimizing Information Quality in Mobile Crowdsensing Systems

FRANCESCO RESTUCCIA, Northeastern University, USA
PIERLUCA FERRARO, University of Palermo, Italy
TIMOTHY S. SANDERS, Tradebot Inc., USA
SIMONE SILVESTRI, University of Kentucky, USA
SAJAL K. DAS, Missouri S&T, USA
GIUSEPPE LO RE, University of Palermo, Italy

Mobile crowdsensing allows data collection at a scale and pace that was once impossible. One of the biggest challenges in mobile crowdsensing is that participants may exhibit malicious or unreliable behavior. Therefore, it becomes imperative to design algorithms to accurately classify between reliable and unreliable sensing reports. To this end, we propose a novel Framework for optimizing Information Reliability in Smartphone-based participaTory sensing (FIRST), that leverages mobile trusted participants (MTPs) to securely assess the reliability of sensing reports. FIRST models and solves the challenging problem of determining before deployment the minimum number of MTPs to be used in order to achieve desired classification accuracy. We extensively evaluate FIRST through an implementation in iOS and Android of a room occupancy monitoring system, and through simulations with real-world mobility traces. Experimental results demonstrate that FIRST reduces significantly the impact of three security attacks (i.e., corruption, on/off, and collusion), by achieving a classification accuracy of almost 80% in the considered scenarios. Finally, we discuss our ongoing research efforts to test the performance of FIRST as part of the National Map Corps project.

## 1 INTRODUCTION

The past few years have witnessed the proliferation of smartphones in people's daily lives; interestingly, today's smartphones are equipped with a set of cheap but powerful embedded sensors, such

Author's addresses: F. Restuccia, Northeastern University, Boston, MA 02215 USA. Email: frestuc@ece.neu.edu; P. Ferraro and G. Lo Re, University of Palermo, Palermo PA 90128 Italy. Email: {pierluca.ferraro, giuseppe.lore}@unipa.it; T.S. Sanders, Tradebot Inc, Kansas City, MO 64116 USA. Email: sanders@tradebotsystems.com; S. Silvestri, University of Kentucky, Lexington, KY 40506 USA. Email: silvestri@cs.uky.edu; S.K. Das, Missouri University of Science and Technology, Rolla, MO 65401 USA. Email: sdas@mst.edu.

as accelerometer, digital compass, gyroscope, GPS, microphone, and camera. The pervasiveness of smartphones coupled with a near-ubiquitous wireless network infrastructure can thus be leveraged to sense, collect, and analyze data far beyond the scale of what was once possible, without the need to deploy thousands of static sensors. This new paradigm is commonly referred to as *mobile crowdsensing* (MCS) [18]. Realizing its great potential, many researchers have developed various applications based on this paradigm, including road traffic monitoring [13, 14, 46], information sharing [6, 17], environmental and crime monitoring [4, 5, 39], just to name a few.

**Motivations and Challenges.** The inherent collaborative nature of MCS implies that its success is strictly dependent on the *reliability* of the information sent by the participants. On the other hand, it is well recognized that participants may *voluntarily* submit unreliable information. For instance, participants may be maliciously aimed at *degrading* the received service to the other users of the application by conducting security attacks. In March 2014, to give an example, students from Technion-Israel Institute of Technology successfully simulated through GPS spoofing a traffic jam on *Waze* that lasted hours, causing thousands of motorists to deviate from their routes [3]. These (and similar) attacks are made extremely easy by smartphone applications (apps) like *LocationHolic* or *FakeLocation* [28], which allow participants to spoof their current GPS location. A second reason to submit false information is to obtain an *unfair advantage* with respect to other users (e.g., rewards obtained without actually doing any sensing service) [35]. In other words, such unreliable behavior may lead to the degradation of the information quality (IQ) contained in sensing reports.

Table 1 summarizes a list of apps in which malicious participants may obtain an advantage by submitting unreliable information.

| **Application** | **Type** | **Unfair Advantages** |
|---|---|---|
| Foursquare | Location Review | Badges and points received without actually checking-in at locations. |
| Ingress | Social Game | Getting score and level unfairly claiming not visited locations. |
| Shopkick | Store Review | Received rewards for reviews of stores not visited. |
| Uber | Car Trip Finder | Drivers can increase the odd of being called faking their location in points of interest (airport, train station, etc.) |
| Waze | Traffic Monitoring | Getting points through fake travels. |

Table 1. Summary of unfair advantages in MCS apps.

Recognizing the crucial importance of addressing the issue of improving the IQ level in mobile crowdsensing, the research community has recently proposed a number of solutions to address this issue [20, 29, 31, 40, 44, 45]. Most of prior work improves the IQ by designing *classification algorithms* able to accurately discriminate between unreliable and reliable sensing reports. To help in this process, the most popular strategy is to use reputation and trust frameworks [7], which estimate the reliability of sensing reports by keeping track of participants' reliability over time. The reader may refer to our recent survey paper [41] for more information on the topic of IQ in MCS and how it has been tackled in prior work. The issue with existing solutions is that they strongly rely on easy-to-manipulate factors to update the reputation scores, such as participants' location or mobility pattern [31, 44], or consensus-based techniques such as majority voting [20]. Thus, these approaches may not be effective when a significant number of participants are malicious (i.e., they deliberately submit false information).

To ease the impact of malicious and/or unreliable participants, a limited number of *mobile trusted participants* (MTPs) may be employed to help build reputation scores in a secure manner, and thus 'bootstrap' the trust in the system [7]. Specifically, MTPs are participants that are hired by the sensing application to periodically generate reliable reports that reflect the actual status of the event that is being monitored around their location. This methodology is being successfully used in the *National Map Corps* project [32] developed by the U.S. Department of Geographical Survey (USGS), where MTPs (in this case, USGS employees) are employed to validate crowdsourced data, such as the exact location of schools and cemeteries (website at http://nationalmap.gov/TheNationalMapCorps/). MTPs are also used in the *Crowd Sourcing Rangeland Conditions* project [23], where Kenyan pastoralists are recruited as MTPs by researchers to validate sensed data regarding local vegetation conditions. The advantage given by MTPs is the capability to tackle malicious and unreliable behavior by building *reliable* reputation scores. However, MTPs also inevitably represent an additional *cost* for the MCS system, as MTPs need to be recruited.

In this paper, we aim at answering the following questions: *what is the minimum number of MTPs we need to employ to ensure that the system will achieve a certain classification accuracy? How does the mobility of the MTPs affect the optimum number of MTPs needed? What is the impact of non-trivial security attacks, such as on/off and collusion, on the classification accuracy?*

There are several issues that need to be tackled to answer these questions. For example, formalizing the relationship between the number of MTPs deployed and the resulting classification accuracy is significantly complex, since the latter is heavily influenced by the mobility of MTPs and other users. Formalizing such relationship *before the deployment of the MCS system* is even more challenging, due to the fact that mobility information may often be unknown *a priori*.

**Contributions.** These questions motivated our work and the following novel contributions.

(1) We mathematically formulate the *MTP Optimization Problem* (MOP), which aims at minimizing the number of MTPs deployed (to minimize hiring costs) while guaranteeing the desired accuracy in classifying the collected reports as reliable or unreliable;

(2) We propose a novel *Framework to optimize Information Reliability in Smartphone-based participaTory sensing* (FIRST), which has three main components. A probabilistic model, called *Computation of Validation Probability* (CVP), calculates the probability that a user report undergoes validation as a function of the number of MTPs deployed and user mobility. A novel image processing algorithm, named *Likelihood Estimation Algorithm* (LEA), leverages geographical constraints of the sensing area to provide an approximation of the probability that a sensing report will be validated. Finally, an *optimization algorithm* (MOA) efficiently solves the MOP by using the results from CVP and LEA, and computes the minimum number of MTPs to achieve desired classification accuracy;

(3) We extensively evaluate the performance of FIRST by considering a mobile crowdsensing (MCS) application for monitoring road traffic implemented with real mobility traces [2, 38, 49]. For comparison purpose, we also implemented [20, 40]. To test their performance, we consider three security attacks, namely corruption, on/off, and collusion attack [35]. Our experimental results demonstrate that FIRST outperforms the state of the art and achieves high classification accuracy, and is able to tackle effectively all the three attacks considered in this paper;

(4) We further evaluate the performance of FIRST on a practical implementation of a mobile crowdsensing (MCS) system, which was conducted at the IEEE PerCom 2015 conference. In this experiment, we designed an app (for both iOS and Android devices), which was distributed to the interested participants (i.e., volunteers) at the conference. These volunteers sent reports regarding the conference participation, acting as users of the MCS system. Results show that FIRST outperforms previous approaches [20, 40] and achieves on the average a high classification accuracy of about 80%, which accounts for an improvement of about 20%.

**Paper Organization.** This paper is organized as follows. Related work is summarized in Section 2, while Section 3 introduces the system architecture and describes the concept of MTP. Section 4 presents the proposed FIRST framework and its three components. Section 5 presents the experimental results, while Section 6 draws conclusions and discusses ongoing work in collaboration with the USGS.

## 2 RELATED WORK

Recently, the information quality (IQ) problem in mobile crowdsensing (MCS) systems has attracted a tremendous attention from the research community, and is expected to gain additional momentum as MCS systems become more and more pervasive. We refer the reader to our recent paper [41], in which we extensively survey existing work and research challenges in this field.

Related work on IQ in MCS can be divided into two main approaches: *trusted platform modules* (TPMs) and *reputation-based systems*. TPMs are hardware chips that reside on the participants' devices, and ensure that the sensed data is captured by authentic and authorized sensor devices within the system [12, 16, 42]. The main drawback of this approach is that TPMs require additional hardware not currently available on off-the-shelf devices, implying such solutions are not readily deployable. Moreover, TPM chips are tailored to verify data coming from physical sensors (e.g., accelerometer, camera). Thus, they are not applicable to MCS systems in which the information is directly supplied by the participants.

Most of related work has focused on developing reputation-based systems to increase information reliability [35]. Specifically, they associate each user with a *reputation level*, which is estimated and updated over time. Among related work, [20, 29, 40, 44] are the most relevant. In [44], the authors proposed *ARTsense*, a reputation-based framework that includes a privacy-preserving provenance model, a data trust assessment scheme, and an anonymous reputation management protocol. The main issue of [44] is that user reputation is updated by considering contextual factors, such as location and time constraints. Given user location and timestamp of reports are easily forgeable quantities, the solution proposed in [44] may not perform well in practical MCS systems, where malicious users may voluntarily tamper with their GPS location and timestamp of reports. Recently, in [20] the authors proposed a reputation framework which implements an improved version of majority vote. The main limitations of this framework are (i) the assumption of constant sampling rate, which is not realistic in asynchronous MCS systems, and (ii) the poor resilience to a large number of malicious users, as the framework uses a modified version of majority vote to update user reputation levels. To overcome such limitation, in [40] the authors proposed *FIDES*, a reputation-based framework that used mobile security agents (concept that inspired our MTPs in this paper) to classify sensing reports. Similarly to us, FIDES is also resilient to a large number of malicious users. However, as in [20], the necessity to set a significant number of parameters makes the actual performance of the framework hardly predictable in reality. On the other hand, FIRST does *not* depend on specific parameters.

A number of frameworks aimed to recruit participants in order to maximize the coverage of the sensing area have been recently proposed [25, 27, 43, 47, 50, 51]. In [25], the authors propose a framework to ensure coverage of the collected data, localization of the participating smartphones, and overall energy efficiency of the data collection process. Zhang *et al.* proposed in [50] *CrowdRecruiter*, a framework that minimizes incentive payments by selecting a small number of participants while still satisfying probabilistic coverage constraint. In [47], Xiong *et al.* proposed a framework aimed to maximize the coverage quality of the sensing task while satisfying the incentive budget constraint. In [43], the authors formulate the problem of sensing given points of interest as a gamification problem, and devise a heuristic algorithm for deriving the set of users to which requests are sent and appropriate reward points for each request. Our approach is different because we rely on MTPs to compute the trustworthiness of participants, which ultimately improves information quality significantly.

## 3 PRELIMINARIES AND BACKGROUND

In this paper, we consider a mobile crowdsensing architecture (depicted in Figure 1) consisting of a platform (MCSP) which can be accessed through 3G/4G or WiFi Internet connection. The data collection process is as follows. First, participants download through common app markets like *Google Play* or *App Store* the mobile crowdsensing app, which is responsible for handling data acquisition, transmission, and visualization (step 1). Then, the MCSP sends (periodically or when necessary) sensing requests through the cloud to registered participants (step 2). The participants can answer such requests by submitting their sensed data (step 3), and eventually receive a reward for their services (step 4). Hereafter, we will use the words "participant" and "user" interchangeably.

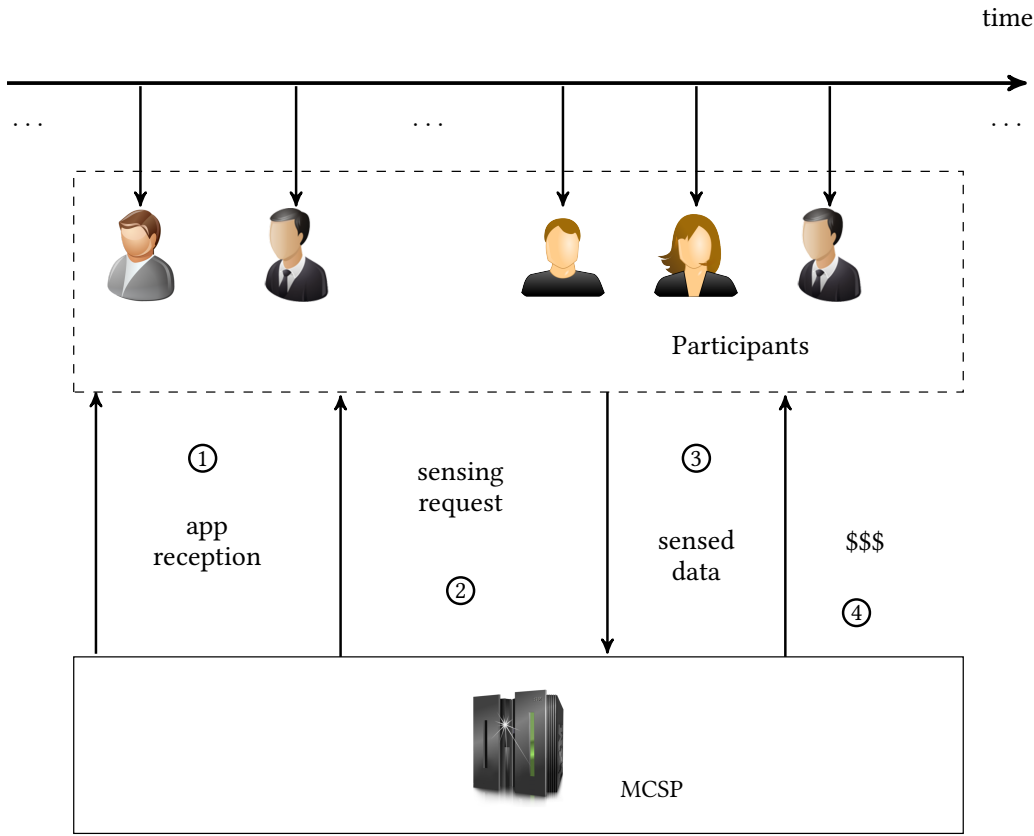

Fig. 1. System architecture.

As far as the sensing application is concerned, we consider a sensing system in which the phenomenon being monitored is **(i)** quantifiable, **(ii)** dynamic (i.e., varies over time), and **(iii)** not subject to personal opinion. This includes phenomena measurable with physical sensors, for example, air/noise pollution levels [10], but also quantities such as occupancy level of parking lots [36], gas prices [11], traffic events (e.g., car crashes and traffic jams) [46], and so on. Furthermore, we assume that the *range* of the sensing quantity being monitored may be divided up into *intervals* or *categories*, which are specific to the MCS application but are properly defined before deployment.

For example, in a traffic monitoring application, a different category for each traffic event (e.g., "Car Crash", "Road Closure", "Traffic Jam", and so on), like in the Waze app [46], could be specified. We define a sensing report as *reliable* if the quantity being reported falls into the interval the phenomenon is currently in (or belongs to that category).

As far as the security assumptions are concerned, we consider the MCSP trustworthy in terms of its functionality (such as user registration, issuing credentials, receiving, processing, and re-distributing data). Furthermore, confidentiality, integrity, and non-repudiation are assumed to be addressed by using standard techniques such as cryptography and digital signatures.

In the following, we concentrate on tackling the inappropriate behavior of participants, and assume they may exhibit *malicious* or *unreliable* behavior; below, we define in details such behavior models. In the following, we will assume users are identified by the MCSP by username and password and some sort of user-unique information (e.g., credit card information), meaning no sybil/rejoin attacks are possible.

- *Malicious:* These users are willingly interested in feeding unreliable reports to the system; their purpose is to either creating a disservice to other users (e.g., fake road traffic lines [3]), or gaining an unfair advantage w.r.t. other users.
- *Unreliable:* These users are not willingly submitting false information, but they still do it because of malfunctioning sensors or incapability in performing the sensing task [40].

FIRST provides a general approach to determine the reliability of each user depending on his/her behavior. We experimentally study in Section 5 three types of attacks, namely the corruption, on-off and collusion attacks (previously defined in [35]), and prove that FIRST is able to quickly detect the malicious behavior and discard unreliable reports. We would like to point out that hereafter we will focus only on the issue of information reliability. Other threats, for example DoS-based attacks, are out of the scope of this paper. Also, note that incentivizing users' participation is out of the scope of this paper; solutions such as [48] may be integrated.

### 3.1 Mobile Trusted Participants

In this paper, we take the same approach used by the successful *National Map Corps* [32] and *Crowd Sourcing Rangeland Conditions* [23] projects, and use mobile trusted participants (MTPs) to tackle the attacks described in the previous section. Specifically, MTPs are individuals who are able and willing to submit regularly reliable reports regarding the phenomenon being monitored or observed. These reports are used to *validate* users' sensing reports coming from nearby, and ultimately estimate the reliability of those participants. Such estimate is used to classify reports generated where MTPs are currently not present, as explained in the next sections.

To allow mathematical formulation, we logically divide up the sensing area into $S = \{s_1, \ldots, s_n\}$ sectors, which may have variable size and represent the *sensing granularity* of the application. For example, in the gas price app, we can have one sector for each gas station. In an air pollution monitoring application, a sector may be as large as a neighborhood of a city, whereas in a traffic monitoring application, sectors may be as large as a city block. We also define $\mathcal{U} = \{u_1, \ldots, u_z\}$ as the set of users contributing to the sensing application.

We model the MTP report validation process as follows. In order to validate user reports, we assume that the reports sent by MTPs are valid for a time period of $T$ units. The value of $T$ is a system parameter that is dependent on the variance over time of the sensing quantity being measured. For example, in a traffic monitoring application, a good value of $T$ could be 5-10 minutes, while in a gas price monitoring app $T$ can be much longer (in Section 5.2, we evaluate the impact of $T$ on the system performance).

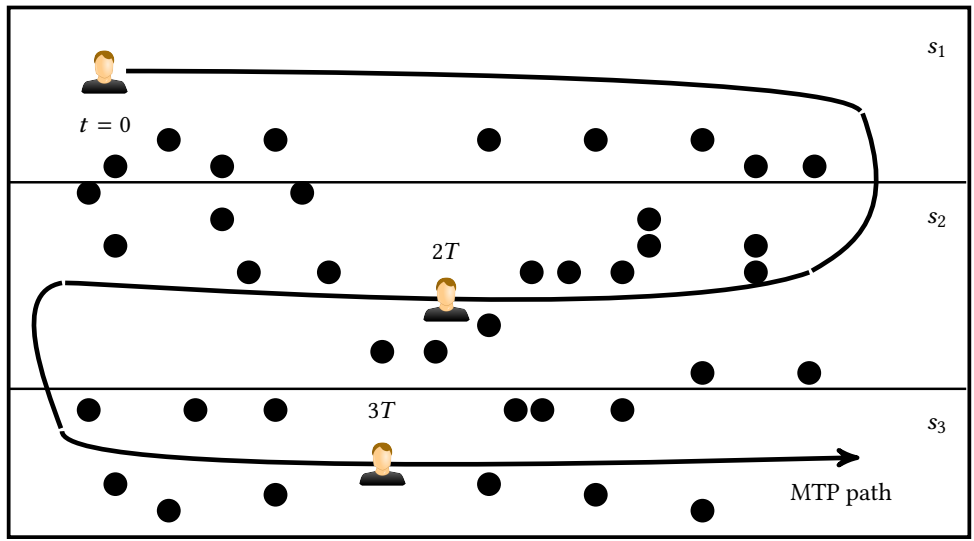

Fig. 2. An MTP moving over the sensing area.

DEFINITION 1. **Validation of sensing reports**. *Whenever a sensing report $q$ is received from a user $u_i$ in sector $s_j$, the platform checks whether a report from an MTP in sector $s_j$ was received in the previous $T$ time units. If yes, then the report is cross-checked with that coming from the MTP. If $q$ is reliable (i.e., falls into the range of the report sent by the MTP), $q$ is marked as* validated *and classified as reliable. Instead, $q$ is rejected if unreliable. If $q$ is not validated, it is classified reliable or unreliable depending on an algorithm discussed in Section 4.*

Figure 2 illustrates an example in which an MTP is moving over a sensing area comprising three sectors. The locations at which the MTP submits a sensing report are marked by a human figure, while users are depicted as black dots. The user reports from sector $s_1$ between $t = 0$ and $T$ units are validated by using the MTP report sent at $t = 0$. Meanwhile, the MTP moves to sector $s_2$ and generates a new report at time $2T$, which then validates users reports from sector $s_2$ in the next time window. Similarly, the MTP report at $3T$ validates the user reports from sector $s_3$ in the time interval $[2T, 3T]$.

Examples of MTPs in urban sensing scenarios include, but are not limited to, professional drivers (i.e., taxi/bus), policemen, employees of the MCS application, or people commuting on a daily basis to their workplace. Henceforth, we will consider the MTPs as reliable, in sense that it is implied that their reports reflects the actual status of the event being monitored. This also implies that reports originating from the same sector during the same time window are supposed to be equivalent. The case in which trusted participants can be (up to some extent) unreliable has already been studied [40]. Since we believe that assuming perfectly controllable MTP mobility may not be realistic in real-world urban sensing scenarios, hereafter we will assume the MTP mobility as not controllable.

### 3.2 MTP Optimization Problem

It is intuitive that the number of validated sensing reports (and therefore, information reliability) increases as the number of recruited MTPs increases. However, in practical implementations, we cannot assume unconstrained budget to recruit MTPs; the number of MTPs that can be used by the system will be limited and therefore, insufficient to guarantee perfect information reliability. To this end, we define the MTP Optimization Problem (MOP). Before that, we define the metric of classification accuracy. Table 2 summarizes the main symbols used in the following mathematical analysis.

| Symbol | Description |
|---|---|
| $\overline{X}$ | Complement of event $X$ |
| $R/\overline{R}$ | Event of the system considering a report as reliable/unreliable |
| $E$ | Event of classification error |
| $V/\overline{V}$ | Event of report validation/non-validation |
| $\mathbb{P}\{E\}$ | Classification error |
| $F/\overline{F}$ | Event of a report being unreliable/reliable |
| $\epsilon^{max}$ | Desired maximum classification error probability |
| $Q$ | Set of MTPs |
| $\mathcal{U}$ | Set of users/participants |
| $\mathcal{S}$ | Set of sectors |
| $u(i; t, z)$ | Probability of user $i$ being in sector $z$ at time $t$ |
| $q(i; t, z)$ | Probability of MTP $i$ being in sector $z$ at time $t$ |

Table 2. Summary of main symbols.

Definition 2. **Classification Error.** *Let $R$ define the event of the system considering a report as reliable, and let $F$ define the event of a user submitting an unreliable sensing report. Let $E$ define the event of erroneously deeming reliable (resp. unreliable) an unreliable (resp. reliable) report. By definition, it follows that the probability of event $E$, denoted $\mathbb{P}\{E\}$, can be computed as*

$$\mathbb{P}\{E\} = \mathbb{P}\{F\} \cdot \mathbb{P}\{R \mid F\} + \mathbb{P}\{\overline{F}\} \cdot \mathbb{P}\{\overline{R} \mid \overline{F}\} \qquad (1)$$

*where $\overline{X}$ is defined as the complement of event $X$. Thus, $1 - \mathbb{P}\{E\}$ represents the **classification accuracy** of the MCS system.*

Let $\epsilon^{max}$ be the desired maximum classification error probability that the MCS system is able to tolerate. The MTP optimization problem (MOP) is then defined as follows.

Definition 3. **MTP Optimization Problem.**

$$\text{Minimize } m \text{ such that } \mathbb{P}\{E\} \leq \epsilon^{max} \qquad \square$$

**Discussions.** In Section 4.3, we discuss how the $\mathbb{P}\{F\}$ quantity can be estimated to solve the MOP. The $\mathbb{P}\{F\}$ takes into account that users may sometimes inadvertently send unreliable information while performing the sensing tasks (e.g., sending blurred pictures by mistake), as well as malicious behavior. Note also that $\mathbb{P}\{E\}$ is defined on the reports that have been submitted by participants and have not been validated. Furthermore, having a significant validation probability $\mathbb{P}\{V\}$ does not make the validated reports useless, but instead that may help increase the information quality level of the mobile crowdsensing system. Indeed, although being validated, these sensing reports may differ between each other to some extent. For example, in a gas price monitoring application, the gas prices reported by four users could be $2.10, $2.45, $2.47 and $2.25 but all belonging to the interval [$2, $2.5]. Regardless, a single value has to be obtained from these reports in order to be used by the system and the participants. To this end, a truth discovery algorithm can be used to merge the different data coming from the participants and thus obtain a more reliable result regarding the event being monitored, which ultimately benefits the information quality level of the system.

## 4 THE FIRST FRAMEWORK

In this section, we propose the FIRST framework. Figure 3 illustrates the main components of the framework, defined as follows.

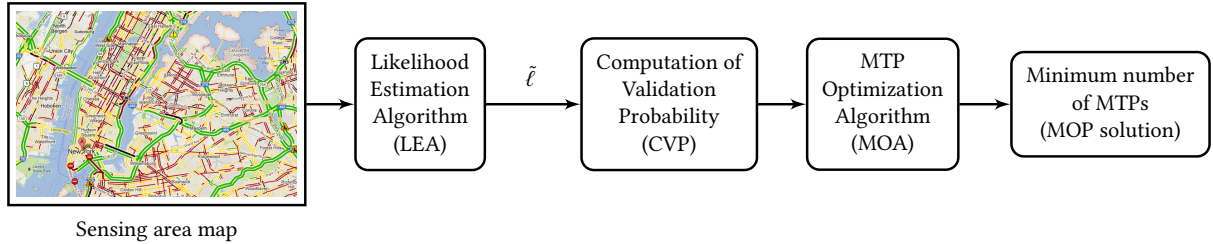

Fig. 3. Block scheme of the FIRST framework.

- **Likelihood Estimation Algorithm** (LEA): It provides an approximation of the mobility of users and MTPs. LEA is based on an *image processing* technique that produces an approximate likelihood based *only* on geographical information (i.e., the map of the sensing area).
- **Computation of Validation Probability** (CVP): This component derives the probability $\mathbb{P}\{V\}$ of the event $V$ that a sensing report will be validated by at least one MTP, as a function of the number of MTPs deployed and the approximate mobility produced by the LEA.
- **MTP Optimization Algorithm** (MOA): It takes $\mathbb{P}\{V\}$ and computes $\mathbb{P}\{E\}$, so as to provide a solution to the MOP to achieve desired maximum error $\epsilon^{max}$.

For better clarity, we first describe the CVP component of FIRST assuming that we have the actual mobility distribution (Section 4.1). Then, we explain how to obtain an approximate distribution of mobility by using the LEA when the mobility is unknown (Section 4.2). Finally, we describe the MOA, and discuss how FIRST is implemented in real-world MCS systems (Section 4.3).

### 4.1 Computation of Validation Probability

In this section, we derive the probability $\mathbb{P}\{V\}$ of the event that a sensing report will be validated by at least one MTP. Let $Q$ be the set of MTPs competing for offering their sensing services, and $\mathcal{U}$ be the set of users of the application. Let $u(i; t, z)$ be the distribution over the sector set $\mathcal{S}$ of the random variable (r.v.) $U_z^t$ describing the location of user $z$ at time $t$. Let also $q(i; t, z)$ be the distribution over the sector set $\mathcal{S}$ of the random variable (r.v.) $Q_z^t$ describing the location of MTP $z$ at time $t$. The notation $x(a; b)$ means that the distribution named $x$ is expressed as a function of the $a$ variable and is parametrized by the variable $b$.

Let us calculate the probability $\mathbb{P}\{V_z \mid U_t^z = s_i\}$ that a sensing report coming from user $u_z$ undergoes validation by an MTP, conditioned to the fact that user $u_z$ is currently in sector $s_i$ of the sensing area:

$$\mathbb{P}\{V_z \mid U_t^z = s_i\} = 1 - \prod_{k \in Q}(1 - q(i; t, k)) \tag{2}$$

In the above equation, we assume that the mobility of each MTP is independent, which is sound because it is highly unlikely MTPs would influence each other's mobility in any way. The above equation can be explained as follows. The probability that a sensing report undergoes validation is the complement of the probability that no MTP is in the same sector as the user. The probability that a sensing report undergoes validation, irrespective of the location of the user, can thus be computed by using the theorem of total probability, i.e.,

$$\mathbb{P}\{V_z\} = \sum_{i=1}^{n} \mathbb{P}\{V_z \mid U_t^z = s_i\} \cdot u(i; t, z) \quad (3)$$

The probability $\mathbb{P}\{V\}$ that on the average a sensing report will be validated can be computed as the average $\mathbb{P}\{V_z\}$ over all the users, which is

$$\mathbb{P}\{V\} = \frac{1}{|\mathcal{U}|} \sum_{z=1}^{|\mathcal{U}|} \mathbb{P}\{V_z\} \quad (4)$$

The probability $\mathbb{P}\{V\}$ that a sensing report will be validated by at least one MTP is assuming that users may have different mobility distributions $\mathbb{P}\{V_z\}$. Therefore, if information regarding the mobility of each user is available, it can be used to compute a more precise estimate of $\mathbb{P}\{V\}$.

**Example.** Figure 4 shows two sensing areas ($S_1$ and $S_2$) divided into the same number $n = 8$ of sectors. We assume that a total of $m = 5$ MTPs are present. For simplicity, in this example we assume that the mobilities of users and MTPs follow the same distribution and that all users follow the same distribution.

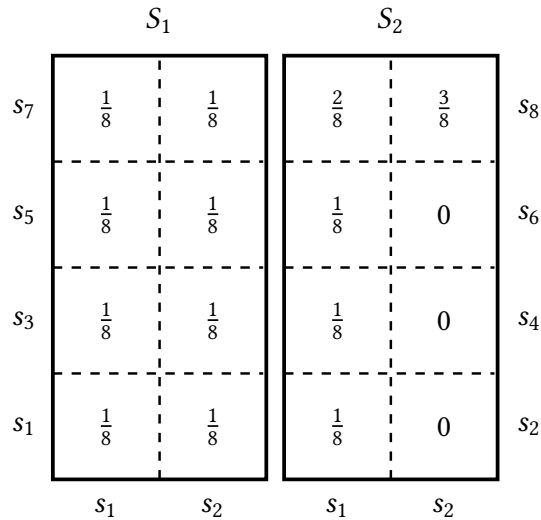

Fig. 4. Example to illustrate computation of $\mathbb{P}\{V\}$.

For simplicity, let us define as $\ell_i^j$ as the probability that an MTP will be in sensing area $j$ and sector $i$. The corresponding mobility distributions $\ell_i^1$ and $\ell_i^2$ are given as: $\ell_i^1 = 1/8$ for $1 \leq i \leq 8$, while

$$\ell_i^2 = \begin{cases} \frac{1}{8} & i = 1, 3, 5 \\ \frac{2}{8} & i = 7 \\ \frac{3}{8} & i = 8 \\ 0 & i = 2, 4, 6 \end{cases} \quad (5)$$

Let us compute $\mathbb{P}\{V\}$ for both sensing areas. First, we need to compute $\mathbb{P}\{V \mid U = s_i\}$ for each $s_i$, which is

- $S_1 : \mathbb{P}\{V \mid U = s_i\} = 1 - (1 - 1/8)^5 = 0.49$ for every $i$, since $\ell_i$ is equal for each sector. Therefore, $\mathbb{P}\{V\} = 1/8 \cdot 8 \cdot 0.49 = 0.49$.

- $S_2: \mathbb{P}\{V \mid U = s_1\} = \mathbb{P}\{V \mid L = s_3\} = \mathbb{P}\{V \mid U = s_5\} = 1 - (1 - 1/8)^5 = 0.49$. $\mathbb{P}\{V \mid U = s_7\} = 1 - (1 - 2/8)^5 = 0.76$, $\mathbb{P}\{V \mid U = s_8\} = 1 - (1 - 3/8)^5 = 0.90$, $\mathbb{P}\{V \mid U = s_2\} = \mathbb{P}\{V \mid U = s_4\} = \mathbb{P}\{V \mid U = s_6\} = 0$. Therefore, $\mathbb{P}\{V\} = 3/8 \cdot 0.49 + 2/8 \cdot 0.76 + 3/8 \cdot 0.90 = 0.71$.

## 4.2 Likelihood Estimation Algorithm

Estimating the mobility distributions $u$ and $q$ is paramount to compute $\mathbb{P}\{V\}$ and therefore, provide a cost-efficient solution to the MOP. In cases where information about the mobility of users and MTPs is available, for example, mobility traces of MTPs and users are available, an exact computation of $u$ and $q$ may be used. However, prior mobility information may not always be available.

In this paper, we developed a heuristic *Likelihood Estimation Algorithm* (LEA) to provide a tighter bound on the mobility of users and MTPs, with just knowing the sensing area location. This heuristic is based on the following rationale: the MCS systems we are considering are deployed in cities, or anyway close to urban areas. This implies that the mobility of users and MTPs will be likely to be almost restricted to the main arterial roads of the sensing areas, or anyway the zones/roads with the greater amount of traffic (both pedestrian and vehicular). By restricting the possible area of movement of the MTPs and users, we are able to reduce the randomness of the movement of users and MTPs, and therefore, provide a tighter bound on the likelihood of sectors.

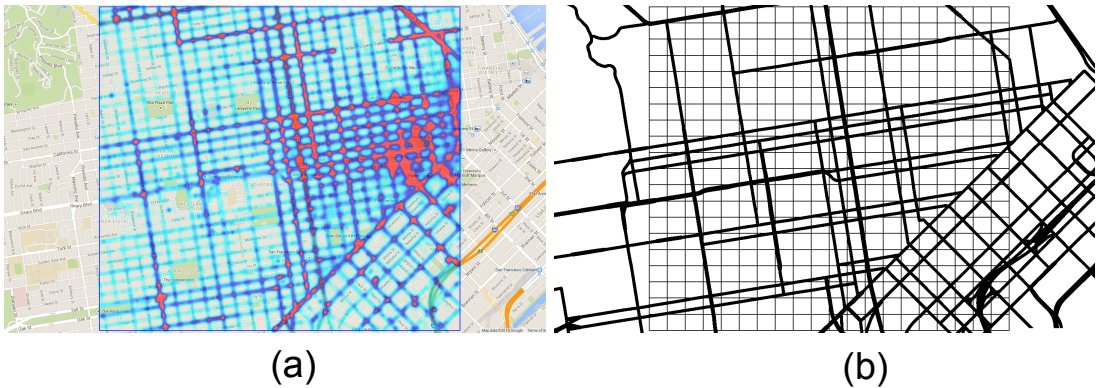

Fig. 5. (a) Heatmap of traces vs. (b) Arterial roads.

Let us now describe the LEA algorithm which works as follows. We consider the map $M$ of the sensing area, and divide it into $n$ sectors as required by the application, where $\mathcal{S} = \{s_1, \cdots, s_n\}$ is the set of sectors. Then, information about the most popular places (which may be roads/squares/buildings) and the geographical constraints of the sensing area is acquired. By using Google Maps APIs, we highlight the main arterial roads on a specific location area. This information is leveraged to mark such places in the map $M$, the background of which is further removed to get a black-and-white image of the sensing area as shown in Figure 5(b), where the black pixels represent the popular places.

The LEA is described by the pseudo-code in Algorithm 1. In Section 5, we show that the LEA is remarkably effective in approximating the mobility distribution of users in various settings, by using real-world mobility traces collected in three major cities in three different continents, namely Rome, San Francisco and Beijing.

**Discussions.** The implicit assumptions that LEA makes are (i) the mobility of users and MTPs is stationary (i.e., does not change over time, thus are dispersal); and (ii) users and MTPs follow the same mobility distributions. We also would like to point out that we have used the Google Maps Traffic application programming interfaces (APIs) [21] as our source of information for roads

---

**ALGORITHM 1:** Likelihood Estimation Algorithm (LEA)

---

**Input:** $M$, map of the sensing area
**Output:** $\tilde{\ell}$, approximate distribution of mobility
1: $\mathcal{S} \leftarrow$ set of sectors $s_1 \cdots s_n$
2: $I \leftarrow$ processed image with most popular areas
3: $B \leftarrow 0$ (sum of black pixels in sensing area)
4: **for** each sector $s_i \in \mathcal{S}$ **do**
5: $\quad B_i \leftarrow$ number of black pixels $\in s_i$
6: $\quad B \leftarrow B + B_i$
7: **end for**
8: **for** each $s_i \in \mathcal{S}$ **do**
9: $\quad \tilde{\ell}(i) \leftarrow B_i/B$
10: **end for**
11: **return** $\tilde{\ell}$

---

popularity. By using this API, it is possible to obtain a map highlighting the most popular roads of a city (called in Google Maps "arterial roads") according to usual volume of traffic. Unfortunately, the APIs of Google Maps are proprietary and thus it is not possible to access individual numeric data regarding the popularity of a particular road or geographical area. Thus, we have designed a heuristic algorithm based on the images provided by Google Maps to approximate the popularity of each road, and thus of each sector. On the other hand, LEA is not tied to a particular map application, and other approaches, such as Open Street Maps [15], could be also used.

Although we recognize that LEA undertakes pretty strong assumptions, in the experimental evaluation conducted in Section 5 we show that LEA provides a pretty good approximation of the likelihood of the sectors, considering that we are using information only from a map. Indeed, we don't claim LEA is a fine-grained mobility estimation algorithm. Instead, it is a simple heuristic that provides before deployment an approximate information regarding the likelihood of certain sectors with respect to others. If more reliable information about the mobility is known, it could be used to complement LEA's analysis and achieve better optimization results. Indeed, another (more accurate) MOP solution could be calculated once the system has been deployed and we have more reliable information on the participants' mobility.

### 4.3 Solving the MTP Optimization Problem

Let us now describe the methodology adopted by FIRST to solve the MTP Optimization Problem (MOP) defined in Section 3.2. The first step is to compute the error probability $\mathbb{P}\{E\}$. This implies we need to derive $\mathbb{P}\{R \mid F\}$ and $\mathbb{P}\{\overline{R} \mid \overline{F}\}$, defined in Equation (1), as a function of $\mathbb{P}\{V\}$.

Since in Equation (4) we have shown how to compute $\mathbb{P}\{V\}$ given $q$ and $u$, we now need to apply probability theory to derive these quantities. The $\mathbb{P}\{E\}$ quantity depends on $\mathbb{P}\{R \mid F\}$ and $\mathbb{P}\{\overline{R} \mid \overline{F}\}$, which in turn depend on the quantities $\mathbb{P}\{R \cap F\}$ and $\mathbb{P}\{R \cap \overline{F}\}$. First, $\mathbb{P}\{R \cap F\}$ is derived through the following steps:

$$\mathbb{P}\{R \cap F\} = \overset{0}{\cancel{\mathbb{P}\{R \cap F \mid V\}}} \cdot \mathbb{P}\{V\} + \mathbb{P}\{R \cap F \mid \overline{V}\} \cdot \mathbb{P}\{\overline{V}\} \tag{6a}$$

$$= \frac{\mathbb{P}\{R \cap F \cap \overline{V}\}}{\cancel{\mathbb{P}\{\overline{V}\}}} \cdot \cancel{\mathbb{P}\{\overline{V}\}} = \tag{6b}$$

$$= \mathbb{P}\{R \cap \overline{V}\} \cdot \mathbb{P}\{F\} = \tag{6c}$$

$$= \mathbb{P}\{R \mid \overline{V}\} \cdot \mathbb{P}\{\overline{V}\} \cdot \mathbb{P}\{F\} \tag{6d}$$

The elimination of the quantity $\mathbb{P}\{R \cap F \mid V\}$ in Equation (6a) is due to the fact that the system does not deem a report reliable when it has been validated and marked as unreliable, thus, the event has probability of occurring equal to zero. Next, the conclusion that $\mathbb{P}\{R \cap F \mid \overline{V}\} = \mathbb{P}\{R \cap F \cap \overline{V}\} \cdot 1/\mathbb{P}\{\overline{V}\}$ in Equation (6b) and $\mathbb{P}\{R \cap \overline{V}\} = \mathbb{P}\{\overline{V}\} \cdot \mathbb{P}\{R \mid \overline{V}\}$ in Equation (6d), respectively, follow the definition of conditional probability [26]. Finally, Equation (6d) follows from the independence of events $R \cap \overline{V}$ and $F$. Specifically, the fact that a report is accepted while not validated, which is, the $R \cap \overline{V}$ event, does not depend on the report being actually reliable or not, which is, the $F$ event. This is due to the fact that the system has no knowledge that the report is actually reliable when making an acceptance or rejection decision. Next, the probability $\mathbb{P}\{R \cap \overline{F}\}$ is derived as follows:

$$\mathbb{P}\{R \cap \overline{F}\} = \overset{\mathbb{P}\{\overline{F}\}}{\cancel{\mathbb{P}\{R \cap \overline{F} \mid V\}}} \cdot \mathbb{P}\{V\} + \mathbb{P}\{R \cap \overline{F} \mid \overline{V}\} \cdot \mathbb{P}\{\overline{V}\} = \tag{7a}$$

$$= \mathbb{P}\{\overline{F}\} \cdot \mathbb{P}\{V\} + \frac{\mathbb{P}\{R \cap \overline{F} \cap \overline{V}\}}{\cancel{\mathbb{P}\{\overline{V}\}}} \cdot \cancel{\mathbb{P}\{\overline{V}\}} = \tag{7b}$$

$$= \mathbb{P}\{\overline{F}\} \cdot \mathbb{P}\{V\} + \mathbb{P}\{R \cap \overline{V}\} \cdot \mathbb{P}\{\overline{F}\} = \tag{7c}$$

$$= \mathbb{P}\{\overline{F}\} \cdot \mathbb{P}\{V\} + \mathbb{P}\{\overline{V}\} \cdot \mathbb{P}\{\overline{F}\} \cdot \mathbb{P}\{R \mid \overline{V}\} = \tag{7d}$$

$$= \mathbb{P}\{\overline{F}\} \cdot (\mathbb{P}\{V\} + \mathbb{P}\{\overline{V}\} \cdot \mathbb{P}\{R \mid \overline{V}\}) \tag{7e}$$

In Equation (7a), it is observed that $\mathbb{P}\{R \cap \overline{F} \mid V\} = \mathbb{P}\{\overline{F}\}$. This is because, if a report has been validated, the acceptance of the report depends only on the fact that the report is reliable or not, which happens with probability $\mathbb{P}\{\overline{F}\}$. Similar to Equation (6), Equations (7b) and (7d) follow from the definition of conditional probability, and Equation (7c) follows from the independence of events $R \cap \overline{V}$ and $\overline{F}$.

We can now derive $\mathbb{P}\{R \mid F\}$ and $\mathbb{P}\{\overline{R} \mid \overline{F}\}$ from Equations (6) and (7), as reported below. $\mathbb{P}\{R \mid F\}$ is derived from $\mathbb{P}\{R \cap F\}$ as follows, by applying the definition of conditional probability:

$$\mathbb{P}\{R \mid F\} = \frac{\mathbb{P}\{R \cap F\}}{\mathbb{P}\{F\}} \tag{8}$$

Next, $\mathbb{P}\{\overline{R} \mid \overline{F}\}$ is derived from $\mathbb{P}\{R \cap \overline{F}\}$ as follows:

$$\mathbb{P}\{\overline{R} \mid \overline{F}\} = 1 - \mathbb{P}\{R \mid \overline{F}\} \tag{9a}$$

$$= 1 - \frac{\mathbb{P}\{R \cap \overline{F}\}}{\mathbb{P}\{\overline{F}\}} \tag{9b}$$

Equation (9a) follows from the fact that the events $(R \mid \overline{F})$ and $(\overline{R} \mid \overline{F})$ are complementary; thus, $\mathbb{P}\{R \mid \overline{F}\} + \mathbb{P}\{\overline{R} \mid \overline{F}\} = 1$. Equation (9b) is derived by applying the definition of conditional probability.

The quantity $\mathbb{P}\{R \mid \overline{V}\}$, which is the probability to deem a report reliable when not validated by MTPs, mathematically models the rationale that FIRST employs when making a decision regarding the reliability/unreliability of a non-validated report. To this end, FIRST leverages validated data from MTPs to infer the "trustworthiness" level of a user, and uses this quantity to infer the reliability of sensing reports coming from that user. The rationale is that, if the user has behaved correctly in the past, i.e., most of her validated reports were reliable, then it is likely that such user will be reliable in the future. FIRST implements this rationale by using the Jøsang's trust model [24]. We used this trust model since it has been widely adopted and accepted as a reliable and effective way to mathematically model trustworthiness in a variety of fields [35]. For the sake of simplicity, as stated in Section 4.2, in the analytical model we have assumed that the users follow the same $\mathbb{P}\{V\}$ and $\mathbb{P}\{F\}$ probability distributions. Consequently, $\mathbb{P}\{R \mid \overline{V}\}$ is computed as

$$\mathbb{P}\{R \mid \overline{V}\} = \overbrace{\mathbb{P}\{V\} \cdot \mathbb{P}\{\overline{F}\}}^{\text{belief component}} + \overbrace{{}^{1}\!/\!_{2} \cdot \mathbb{P}\{\overline{V}\}}^{\text{uncertainty component}} \tag{10}$$

This formula implements the Jøsang's trust model as follows. The first part, $\mathbb{P}\{V\} \cdot \mathbb{P}\{\overline{F}\}$, represents the belief component of the users trustworthiness level – it is higher when users are validated most of the time (i.e., $\mathbb{P}\{V\}$ close to 1) and the reports are reliable. The second part, ${}^{1}\!/\!_{2} \cdot \mathbb{P}\{\overline{V}\}$, represents the uncertainty component of the users trustworthiness level – it is higher when most of the reports have not been validated. Note that, as $\mathbb{P}\{V\}$ increases, the value of $\mathbb{P}\{R \mid \overline{V}\}$ approximates to $\mathbb{P}\{\overline{F}\}$. Also, if $\mathbb{P}\{V\} = 0$, the system deems as reliable every report with probability ${}^{1}\!/\!_{2}$ (coin tossing), which is sound as there is no reason to be more inclined to accept or reject the report if no information is available [22].

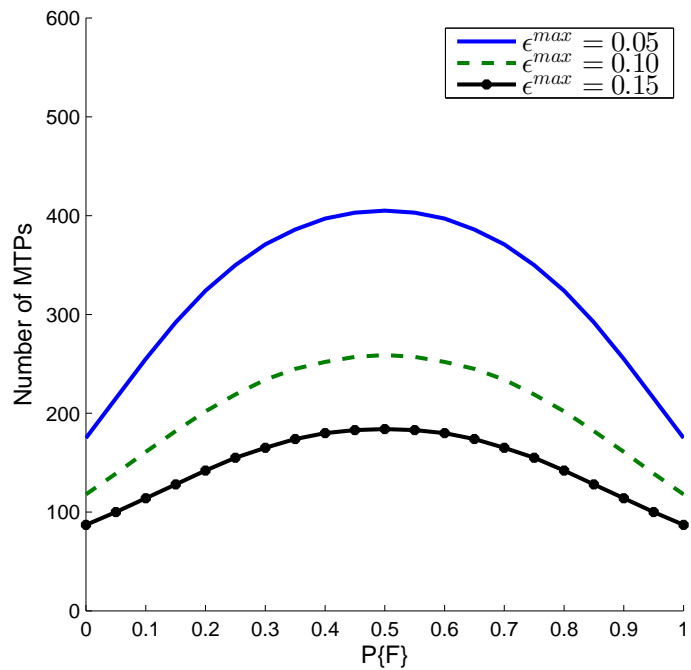

Fig. 6. Number of MTPs vs. $\mathbb{P}\{F\}$.

**Discussions.** An initial value of the probability $\mathbb{P}\{F\}$ is needed in order to solve the MTP optimization problem (MOP). To investigate this aspect, we have run experiments to evaluate the impact of $\mathbb{P}\{F\}$ on the number of MTPs needed to obtain a given classification error. Figure 6 depicts the number of MTPs needed as a function of $\mathbb{P}\{F\}$, for three values of desired maximum error probability $\epsilon^{max}$. Figure 6 shows that, irrespective of $\epsilon^{max}$, the highest number of MTPs is necessary when $\mathbb{P}\{F\} = 0.5$. These results confirm the intuition that the worst case in terms of number of MTPs is when the participants are most unpredictable, i.e., they could send a reliable or unreliable report with the same probability. Therefore, we conclude that $\mathbb{P}\{F\} = 0.5$ is the value that has be chosen to initially run the MOP, as it represents the worst-case scenario. We point out that an *a priori* estimation of $\mathbb{P}\{F\}$ is only used to provide a reasonable approximation (through the MOP) of the number of MTPs that will be needed to achieve the desired classification accuracy. On the other hand, the value of $\mathbb{P}\{F\}$ may change over time (e.g., a participant is malicious and/or unreliable) and it is different for each user. For this reason, after deployment, FIRST estimates the value of $\mathbb{P}\{F\}$ for each participant *on the run*, as explained in Section 4.4. Although the $\mathbb{P}\{F\}$ quantity is assumed to be constant in the analytical model to simplify the mathematical optimization, we study the resiliency of the system to complex attacks like on/off attack and collusion attack in Section 5.1.2, where we also show that the validated data from MTPs can be used to estimate $\mathbb{P}\{F\}$ and thus re-run the MOP to decrease the number of MTPs over time.

**Optimization Algorithm.** We are now ready to present an algorithm to solve the MOP, called the MOP Optimization Algorithm (MOA). The MOA is based on a modified version of binary search algorithm, called Left-most Insertion Point (LMIP). More specifically, LMIP returns the left-most place (i.e., the minimum value) where $\mathbb{P}\{E\}$ can be correctly inserted (and still maintains the sorted order) in the ordered array of the errors corresponding to a particular choice of $m$. This corresponds to the lower (inclusive) bound of the range of elements that are equal to the given value (if any). Note that LMIP can be applied to solve the MOP due to the fact that $\mathbb{P}\{E\}$ is a monotonically decreasing function of $m$. This is because the product component of the quantity $\mathbb{P}\{V_z \mid U_t^z = s_i\}$ in Equation (2) increases with $m$ (i.e., the cardinality of set $Q$ increases). Consequently, more reports are validated by MTPs, which ultimately leads to a lesser $\mathbb{P}\{E\}$.

The MOA takes as input the approximate distribution $\tilde{\ell}_i$ provided by LEA (equal for participants and MTPs), and also $\mathbb{P}\{F\}$, the desired maximum error $\epsilon^{max}$, and the maximum number $m^{max}$ of MTPs available. It provides as output the optimum number $m^*$ of MTPs to be used to achieve the desired maximum error $\epsilon^{max}$.

---

**ALGORITHM 2:** MOP Optimization Algorithm (MOA)

---

**Input:** $\tilde{\ell}_i, \mathbb{P}\{F\}, \epsilon^{max}, m^{max}$
**Output:** $m^*$
1: $\epsilon^{min} \leftarrow \texttt{CalculateError}(\tilde{\ell}_i, \mathbb{P}\{F\}, m^{max})$
2: **if** $\epsilon^{max} < \epsilon^{min}$ **then**
3:     **return** 'infeasible'
4: **end if**
5: **return** $\texttt{LMIP}(\tilde{\ell}_i, \mathbb{P}\{F\}, \epsilon^{max}, 0, m^{max})$

---

Let us calculate the time complexity of the MOA. LMIP is a variation of binary search, therefore its overall complexity will be $O(x \cdot \log m^{max})$, where $x$ is the complexity of $\texttt{CalculateError}$. It requires constant time to compute $\mathbb{P}\{E\}$ using Equation (1) and $n \cdot m \cdot p$ iterations to compute $\mathbb{P}\{V\}$ using Equation (4), where $n$, $m$, and $p$ are the number of sectors, MTPs and participants, respectively. However, when using LEA, we are assuming that users and MTPs follow the same mobility distributions. Thus, Equation (4) can be computed in $\Theta(n)$, and the overall time complexity of MOA is given by $O(n \cdot \log m^{max})$.

**Example 3.** In the example of Figure 7, we assume the $\ell$ distribution equal to $\ell_i^2$ presented in Figure 4, $\mathbb{P}\{F\} = 0.01$, $m^{max} = 8$ and $\epsilon^{max} = 0.1$. In this case, the LMIP will return $m^* = 4$, since it is the left-most element that provides $\mathbb{P}\{E\} \leq 0.1$.

---

**ALGORITHM 3:** Left-most Insertion Point (LMIP)

---

**Input:** $\tilde{\ell}_i, \mathbb{P}\{F\}, \epsilon^{max}, i, j$
**Output:** $m^*$
 1: **if** $j < i$ **then**
 2:     **return** $i$
 3: **end if**
 4: $mid \leftarrow \lfloor (i+j)/2 \rfloor$
 5: **if** CalculateError($\tilde{\ell}_i, \mathbb{P}\{F\}, mid$) $\leq \epsilon^{max}$ **then**
 6:     **return** LMIP($\tilde{\ell}_i, \mathbb{P}\{F\}, \epsilon^{max}, i, mid - 1$)
 7: **else**
 8:     **return** LMIP($\tilde{\ell}_i, \mathbb{P}\{F\}, \epsilon^{max}, mid + 1, j$)
 9: **end if**

---

| $\mathbb{P}\{E\}$ | 0.29 | 0.17 | 0.11 | 0.07 | 0.05 | 0.03 | 0.02 | 0.02 |
| $m$ | 1 | 2 | 3 | 4 | 5 | 6 | 7 | 8 |

Fig. 7. Example of LMIP.

## 4.4 FIRST Classification Algorithm

Algorithm 4 reports in details the procedure that FIRST adopts to classify reports as reliable (R) or unreliable (U). For each user $u_i$, the system keeps track of the number $k_i$ of sensing reports submitted by the user, the number $k_i^v$ of sensing reports validated by an MTP, and the number $k_i^r$ of reports that have been validated as reliable.

---

**ALGORITHM 4:** FIRST Classification Algorithm for Sensing Reports

---

**Input:** Sensing report $q_i$ from user $u_i$
**Output:** Classification result - reliable (R) or unreliable (U)
 1: $k_i \leftarrow 0, k_i^v \leftarrow 0, k_i^r \leftarrow 0 \ \forall i$
 2: **for** each report $q_i$ **do**
 3:     $k_i \leftarrow k_i + 1$
 4:     **if** $q_i$ has been validated **then**
 5:        $k_i^v \leftarrow k_i^v + 1$
 6:        **if** $q_i$ is deemed reliable **then**
 7:           $k_i^r \leftarrow k_i^r + 1$
 8:           **return** $R$
 9:        **else**
10:           **return** $U$
11:        **end if**
12:     **end if**
13:     **if** $q_i$ has not been validated **then**
14:        $T(u_i) = \frac{k_i^r}{k_i} + \frac{1}{2} \cdot \left(1 - \frac{k_i^v}{k_i}\right)$
15:        $\mathbb{P}\{q_i = R \mid T(u_i)\} = T(u_i)$
16:        $\mathbb{P}\{q_i = U \mid T(u_i)\} = 1 - T(u_i)$
17:        **return** $\arg\max_{c \in \{R,U\}} \mathbb{P}\{q_i = c \mid T(u_i)\}$
18:     **end if**
19: **end for**

---

Let us define $T(u_i)$ as the current trustworthiness level of user $u_i$. As soon as a report $q_i$ is received by user $u_i$, the following steps take place. First, the quantity $k_i \leftarrow k_i + 1$. Then, if the report has been validated, $k_i^v \leftarrow k_i^v + 1$. Next, if the report has been validated as reliable by an MTP, (i) $k_i^r \leftarrow k_i^r + 1$; (ii) the classification result is R (line 8). Conversely, if the report has been validated as non-reliable, classification result is U (line 10).

If $q_i$ has not been validated by an MTP, than the following steps take place. The trustworthiness level $T(u_i)$ is updated by using the following equation, which implements the Jøsang's trust model defined in Equation 10:

$$T(u_i) = \frac{k_i^r}{k_i} + \frac{1}{2} \cdot \left(1 - \frac{k_i^v}{k_i}\right) \in [0, 1] \quad (11)$$

Then, FIRST uses a Bayes estimator classifier, defined as $C : [0, 1] \to \{R, U\}$, where $T(u_i)$ is the *prior distribution*. Bayes estimators are probabilistic classifiers [9], the same type of the well-known Naïve Bayes classifier. Probabilistic classifiers define conditional distributions $\mathbb{P}\{Y \mid X\}$, meaning that for a given $x \in X$, they assign probabilities to all $y \in Y$ (and these probabilities sum to one). The "hard" classification (i.e., assigning the input to a specific output class) is done by selecting the class which has the highest probability. Specifically, the output class $C(q_i)$ is computed as follows:

$$C(q_i) = \arg\max_{c \in \{R, U\}} \mathbb{P}\{q_i = c \mid T(u_i)\}, \quad (12)$$

where conditional probabilities are defined as follows: $\mathbb{P}\{q_i = R \mid T(u_i)\} = T(u_i)$ (line 15), and $\mathbb{P}\{q_i = U \mid T(u_i)\} = 1 - T(u_i)$ (line 16). In other words, the trustworthiness level of a user is directly translated to the trustworthiness level of the report. After being classified as reliable, reports may be subsequently analyzed by additional algorithms (for example, [33, 34]) to determine the actual status of the sensing area by combining or fusing the information conveyed by the reliable reports. Note that Algorithm 4 computes the reliability of each report $q_i$ in O(1) time.

## 5 EXPERIMENTAL RESULTS

In this section, we present the experimental results obtained by evaluating the performances of FIRST and comparing it with relevant related work. First, we report the performance results obtained by considering an application monitoring vehicular traffic events. Then, we discuss results obtained by using the Participatory PerCom application.

### 5.1 Participatory Traffic Sensing

To implement this experiment, we considered mobility traces collected from the following datasets:

- *CRAWDAD-SanFrancisco* [38]: This dataset contains mobility traces of approximately 500 taxis in San Francisco, USA, collected over one month's time;
- *CRAWDAD-Rome* [2]: In this dataset, 320 taxi drivers in the center of Rome were monitored during March 2014;
- *MSR-Beijing* [49]: This dataset collected by Microsoft Research Asia contains the GPS positions of 10,357 taxis in Beijing during one month.

In these experiments, we consider a traffic sensing application in which taxi cab drivers report traffic anomalies. We consider sensing areas of approximately 4×4km square areas, which characterize the downtown of cities such as San Francisco, Rome, and Beijing. We implemented the application using the OMNeT++ simulator (available at https://www.omnetpp.org). Anomaly reports are binary (i.e., "there is/there is not a traffic anomaly in a particular sector"), and are generated every five minutes through simulation. The probability of the report being is set to $\mathbb{P}\{F\}$, which varies according to the experiment being run. This allowed us to effectively emulate a road traffic monitoring application. We point out that the following simulations are aimed at validating the performance of the components of the FIRST framework and the resiliency against the *on/off, corruption* and *collusion* attacks. In order to validate the FIRST framework with real-world participants as well as real data, we have also conducted the Participatory PerCom experiment, discussed in Section 5.2.

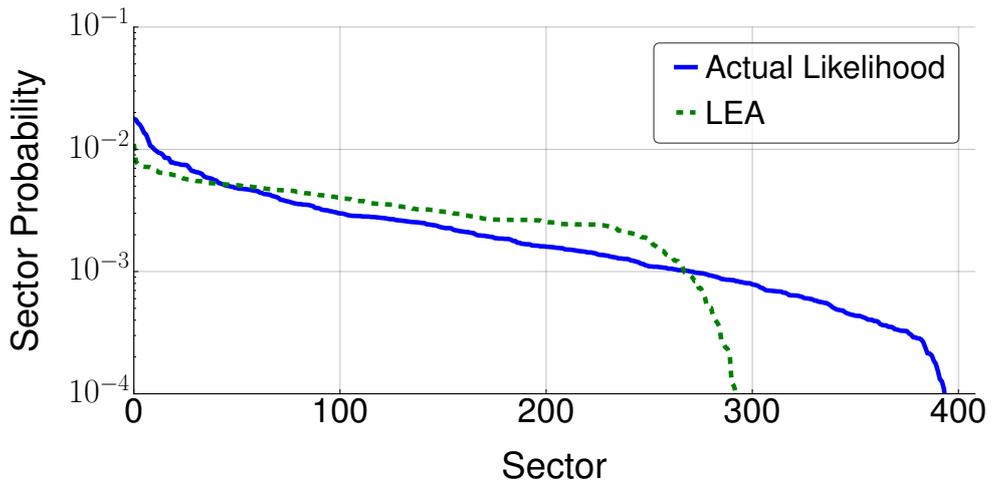

(a) San Francisco

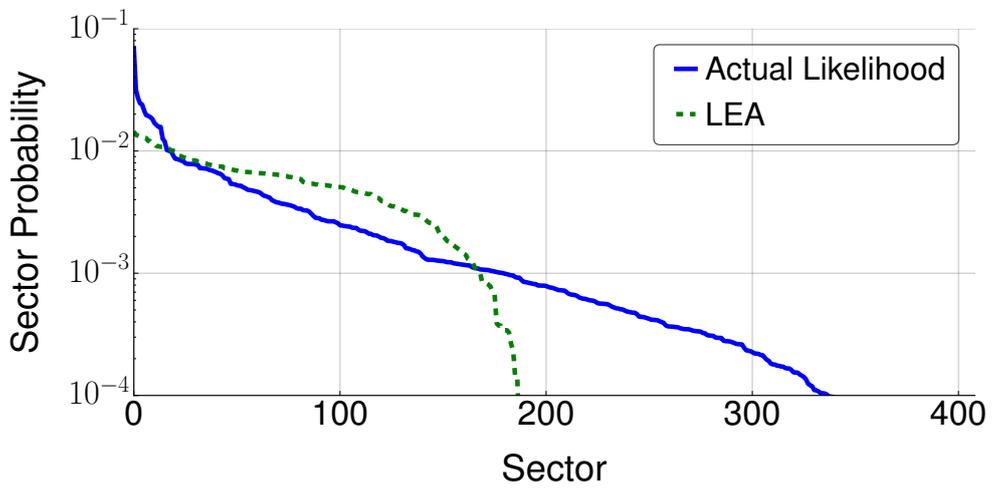

(b) Rome

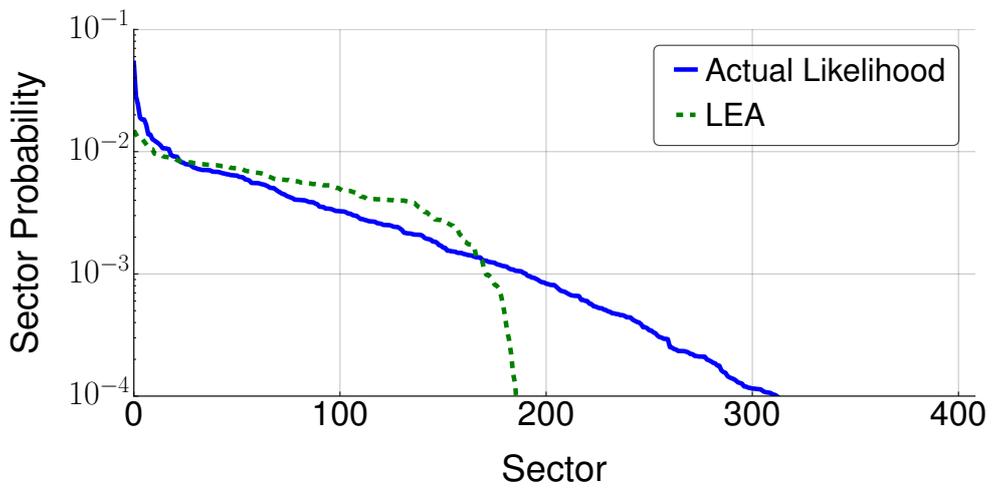

(c) Beijing

Fig. 8. Traces vs. Mobility Estimation Algorithm.

For pre-processing, we have first stored the traces belonging to the three different datasets into the same format. We have also discretized the mobility of the taxi in sectors, according to the topology described in the paper. Furthermore, we have defined a time window size of one minute. At the beginning of each experiment, a portion of taxis is selected as MTPs and another portion as the participants. If we need more taxis than the number of traces (such as in the Rome scenario), we reuse the same traces for differents taxis. The starting point of each mobility trace is chosen randomly for each experiment according to a uniform distribution. The mobility points that are outside the sensing area are discarded.

*5.1.1 Evaluation of FIRST components.* We now evaluate the LEA and MOA components of the FIRST framework. The goal of the first set of experiments is to test the efficacy of LEA in computing the likelihood of sectors. To obtain ground-truth information about the actual mobility of taxi cabs, we processed the traces using OMNeT++. We then ran the LEA algorithm to determine how well it could approximate the true mobility statistics. To apply LEA, we have divided the sensing area into a grid of 20×20 sectors, with sectors having the same size as a city block.

Figure 8 shows the distribution of the likelihood of sectors and the one obtained by LEA, respectively. More specifically, the figure shows the actual and estimated probability of a taxi to be in each sector of the sensing area. Note that, for better clarity, the y-axis in Figure 8 is logarithmic, and the quantities involved are very small (i.e., up to $10^{-4}$). Thus, these experiments conclude that the LEA algorithm approximates well the likelihood of sectors, considering the scarce information available. Figure 9 shows $\mathbb{P}\{E\}$ as a function of the MTPs per sector density, calculated analytically by the Computation of Validation Probability (CVP) component of FIRST. For comparison purposes, we evaluated CVP by providing as input (i) the distribution computed by LEA as applied to each considered sensing area (*CVP-LEA*, represented by a dashed line), and (ii) the uniform mobility distribution (*CVP-Uniform*, represented by a dotted continuous line) as the baseline approximation. We compare such analytical results with the experiments using the traffic datasets.

As shown in Figure 9, in all three scenarios, *CVP-LEA* computes $\mathbb{P}\{E\}$ with remarkable precision. In particular, the maximum difference obtained is 3.47%, achieved in the Rome setting. Furthermore, Figure 9 shows that the accurate estimation of the mobility provided by the LEA translates into an improved prediction accuracy of CVP w.r.t. the uniform distribution, as *CVP-Uniform* yields a maximum difference of 17.02% in the case of Rome setting.

In Figure 10, we apply the MTP Optimization Algorithm (MOA) to analyze the MTPs per sector density that is necessary by FIRST to provide maximum desired error probability $\epsilon^{max}$. Note that the error probability has been experimentally calculated without including the reports coming from MTPs and the reports that have not been validated. Similarly to the experiments shown in Figure 9, we consider users sending unreliable reports with three probability values $\mathbb{P}\{F\} = 0.01, 0.5$ and 0.9. These results highlight that FIRST is remarkably effective in achieving high accuracy with a low number of MTPs. More specifically, it provides on the average 85% of accuracy with an MTPs per sector density of about 32% in case of Rome and Beijing, and 55% in the case of San Francisco.

Note that higher accuracy values require in general a significant number of MTPs, especially when the behavior of participants becomes hardly predictable (i.e., $\mathbb{P}\{F\} = 0.5$) and the mobility is highly entropic (i.e., in San Francisco setting). Somehow surprisingly, Figure 10 also shows that fewer MTPs are needed when $\mathbb{P}\{F\} = 0.9$ than when $\mathbb{P}\{F\} = 0.5$. Intuitively, this is due to the fact

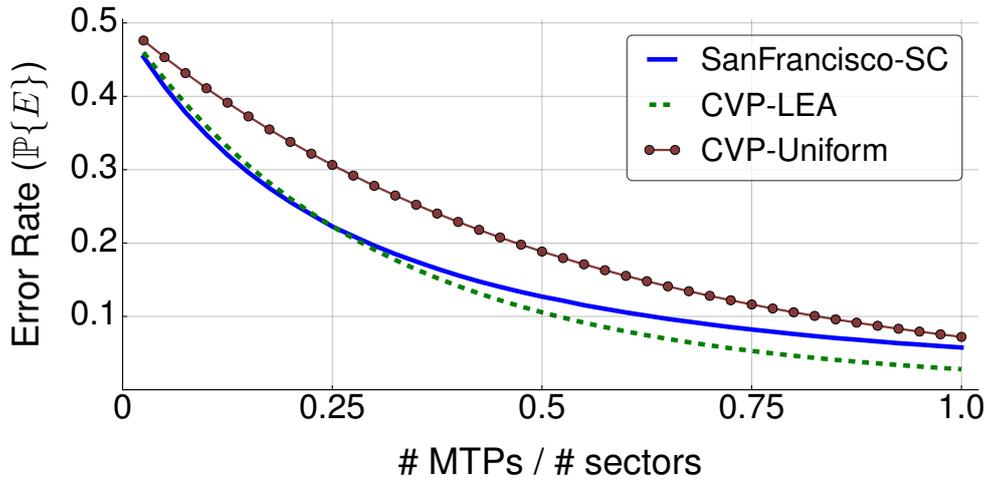

(a) San Francisco

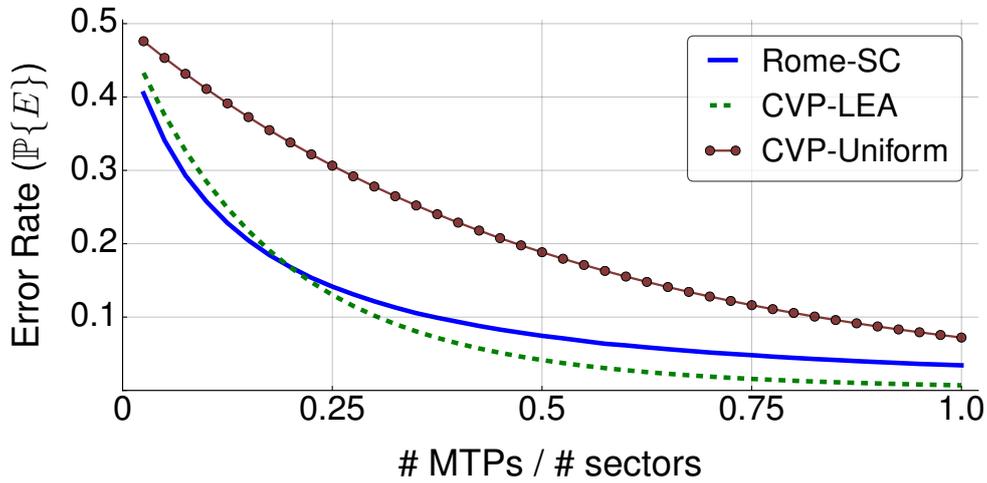

(b) Rome

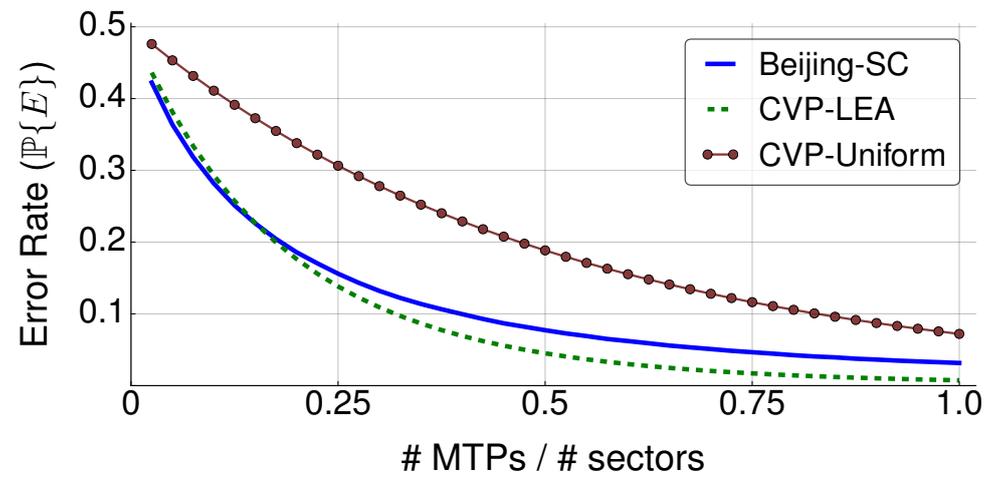

(c) Beijing

Fig. 9. Number of MTPs / number of sectors vs Error Rate ($\mathbb{P}\{E\}$).

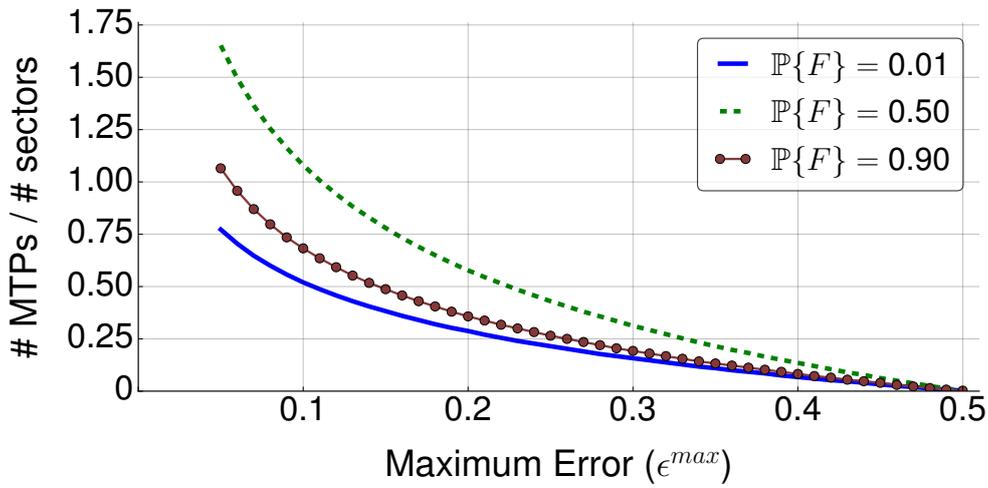

(a) San Francisco

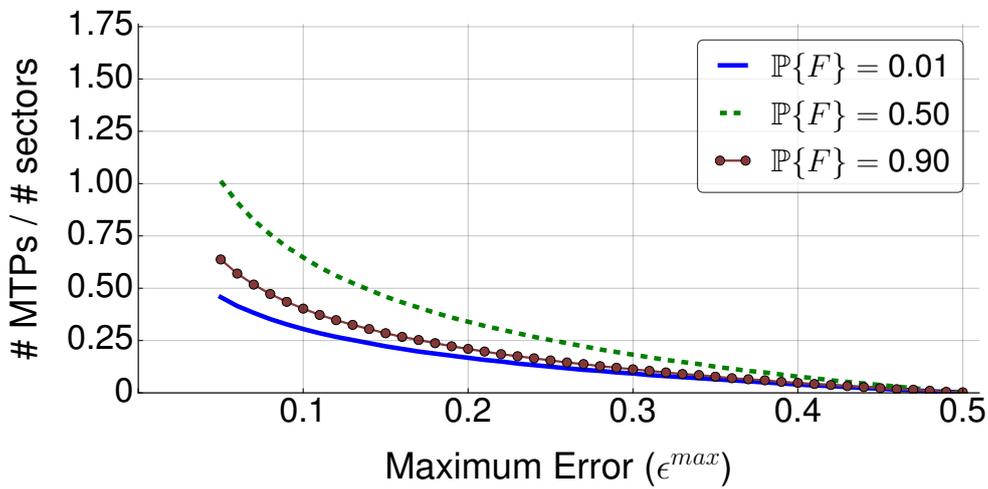

(b) Rome

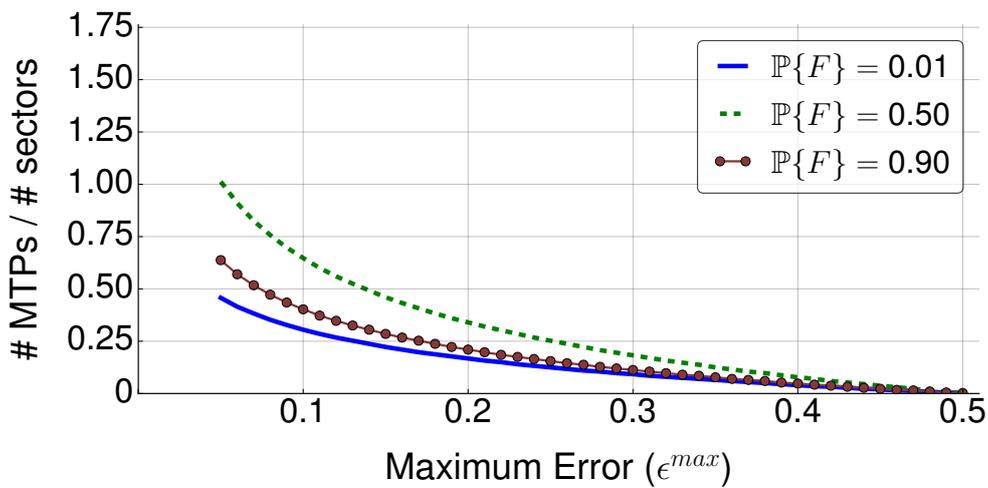

(c) Beijing

Fig. 10. MOA results.

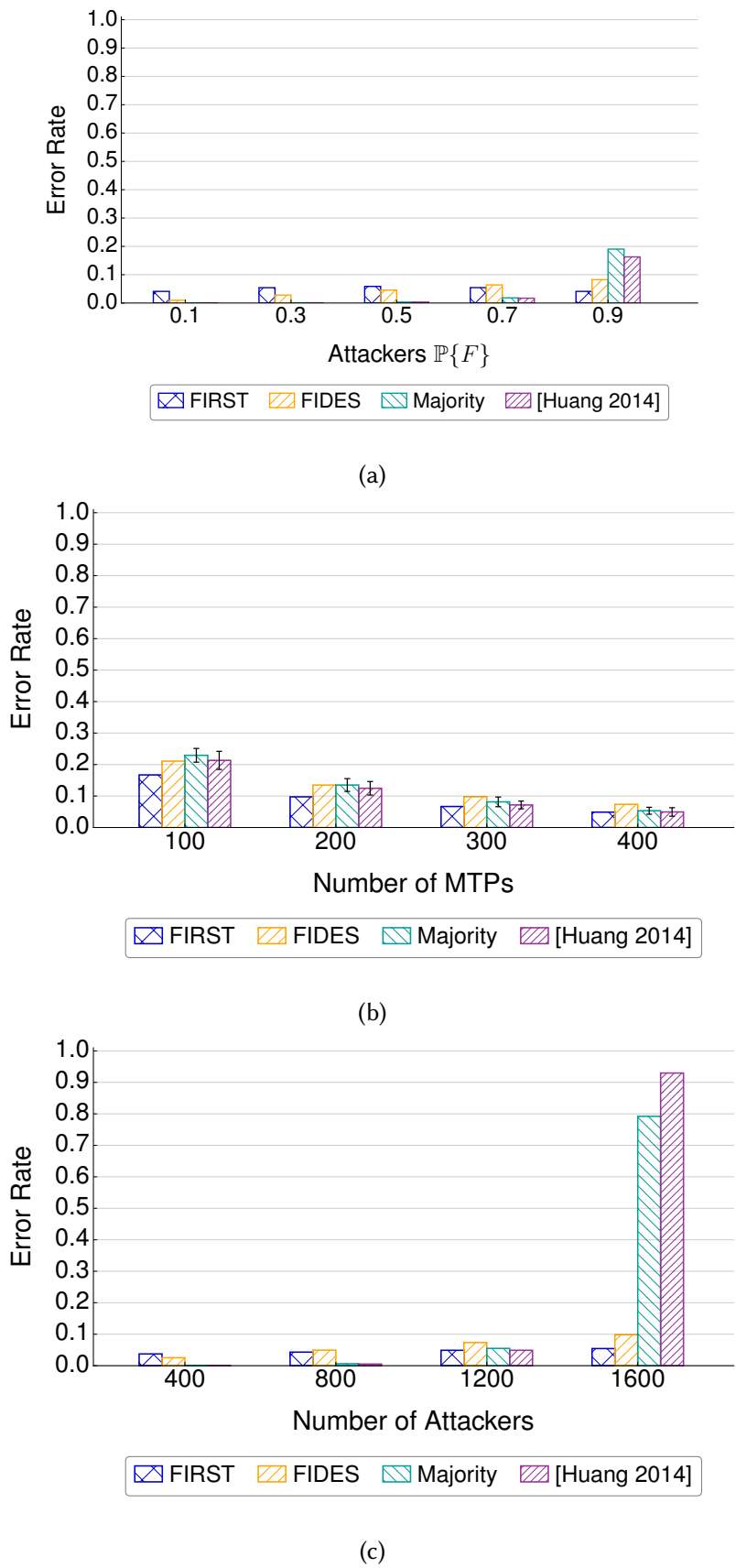

Fig. 11. Corruption attack: Error Rate vs. $\mathbb{P}\{F\}$, MTPs, and attackers.

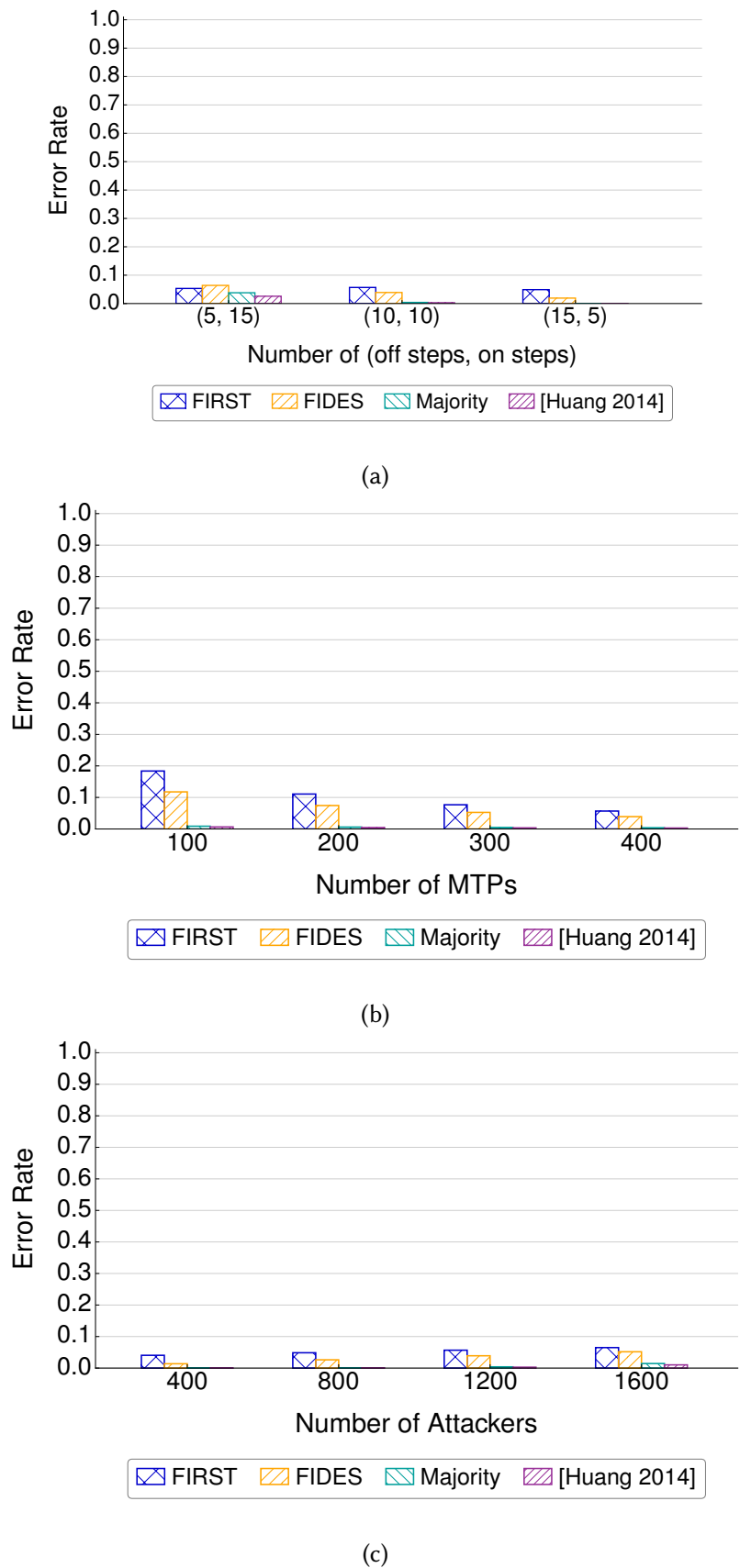

Fig. 12. On-off attack: Error Rate vs. On-off steps, MTPs, and attackers.

that, when participants send reports randomly, it is more difficult to understand their reliability. On the other hand, when their behavior is more "regular" (i.e., consistent over time) it is easier to evaluate their reliability.

From Figures 9 and 10, we also conclude that San Francisco setting requires the most number of of MTPs to achieve the same results obtained in the Rome and Beijing setting. This is because the mobility of MTPs and participants is more concentrated there than in the San Francisco setting. For example, as Figure 10 shows, San Francisco needs approximately 0.5 MTPs per sector to achieve an error probability $\mathbb{P}\{E\} \leq 0.1$ with $\mathbb{P}\{F\} = 0.01$, whereas to achieve the same result in Rome and Beijing approximately 0.3 MTPs per sector are needed.

*5.1.2 Evaluation of attack resiliency.* Based on the behavior models defined in Section 3, in this paper we take into account the following security attacks, which were defined in other domains and recently cast in the context of mobile crowdsensing [35]. For simplicity, hereafter we will generically use the word "attacker" for both malicious and unreliable users, and the words "threat" and "attack" interchangeably.

(1) *Corruption attack.* This threat models the following strategy: for each sensing report, the attacker sends unreliable data with probability $p$ and correct data with probability $1 - p$. This attack can be carried out by unreliable and malicious users alike.
(2) *On-off attack.* In this attack, the malicious user alternates between normal and abnormal behaviors to conceal her maliciousness. Specifically, the adversary *periodically* sends $n$ reliable reports and then $m$ unreliable reports, and then repeats the process. This attack is extremely easy to carry out but also extremely challenging to detect and contrast [1, 8, 37].
(3) *Collusion attack.* In this attack, two or more malicious participants coordinate their behavior in order to provide the same (unreliable) information to the MCSP [19, 30]. The malicious behavior may also include GPS location spoofing, so as to mislead the MCSP into assuming colluding participants are nearby [40].

For comparison reasons, we implemented the FIDES framework [40], and the reputation-based framework proposed in [20], hereafter referred to as [Huang 2014]. FIDES uses a modified version of Jøsang's trust model to update the reputation of users. This framework inherits from Jøsang's trust model a strong sensitivity to parameter tuning. On the other hand, [Huang 2014] proposes an approach which is a improved variation of majority vote, and its performance also depends on the choice of parameter setting (Gompertz's function's, among others). For implementation, we used the parameter settings proposed in the papers, which are reported in Table 3. We also implemented a pure majority vote scheme to obtain baseline performance. Even though Majority and [Huang 2014] do not use MTPs directly, completely ignoring MTP reports for these systems would not be a fair comparison. Thus, MTP reports were considered as normal users' reports when evaluating Majority and [Huang 2014]. If not stated otherwise, in the following experiments we used the parameters reported in Table 3. Confidence intervals at 95% are shown only when above 1% of the value.

Figure 11 reports the error rate (i.e., the percentage of reports erroneously classified as reliable/not reliable) obtained by the frameworks when subject to a corruption attack, as a function of the (constant) attack probability, number of MTPs, and number of attackers. Figure 11(a) and (c) show that the performance of Majority and [Huang 2014] decreases as the number of false reports and attackers increases. This is reasonable, as both schemes are based on data aggregation and therefore not resilient to large number of malicious users and/or unreliable reports. Furthermore, Figure 12 shows the results obtained under the On-off attack by all the considered schemes. As expected, the performance of FIRST is slightly affected by this attack, especially when the percentage of ON steps is less than the OFF one. This is because, the less the ON steps are, the harder it is for FIRST to decrease the accept probability of malicious users. However, FIRST is able to achieve an error rate of about 6% in the worst case. On the other side, [Huang 2014] and Majority are instead more affected when the ON step is greater than the OFF, as it is more likely for them to misclassify sensing reports when the percentage of unreliable reports/number of attackers is higher.

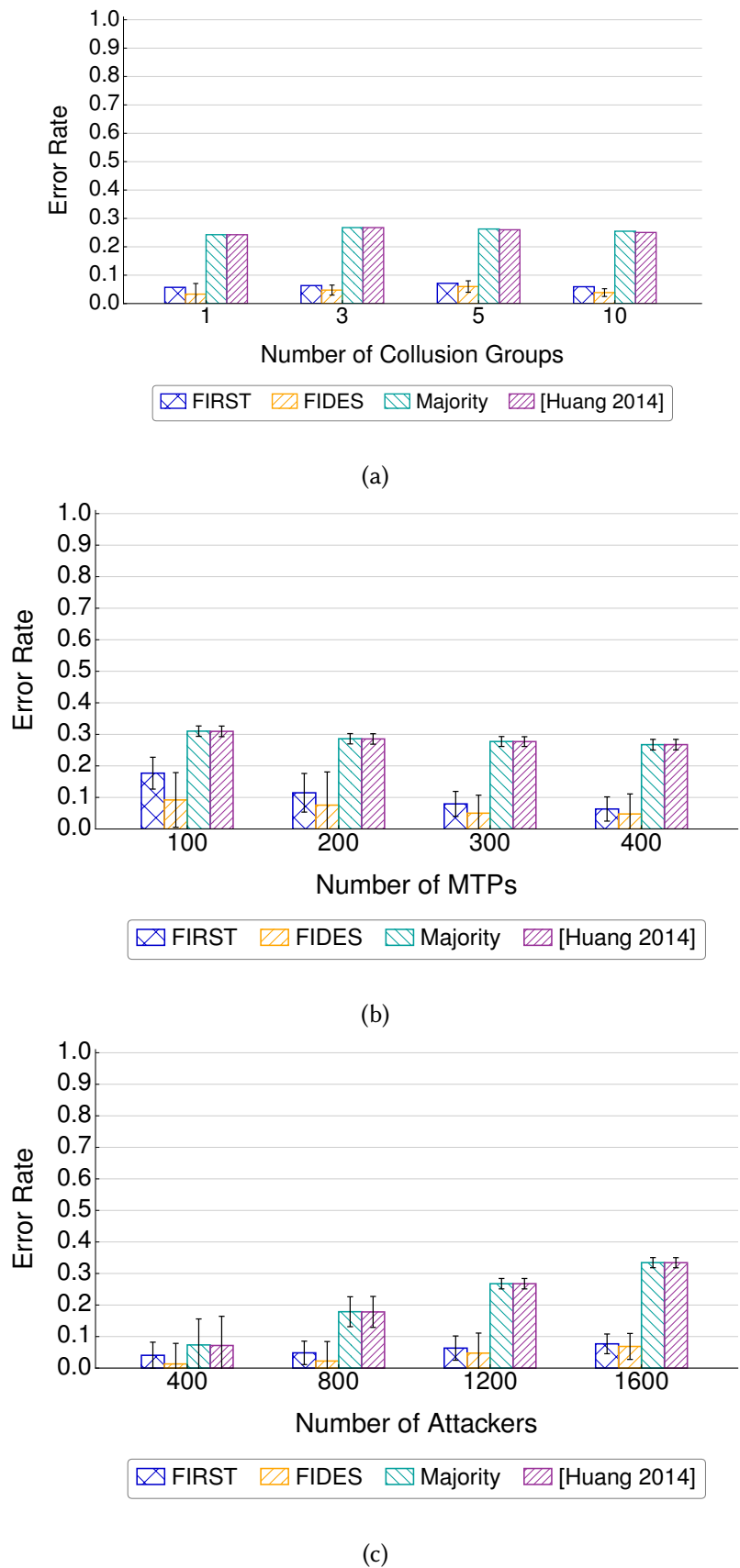

Fig. 13. Collusion attack: Error Rate vs. $\mathbb{P}\{F\}$, MTPs, and attackers.

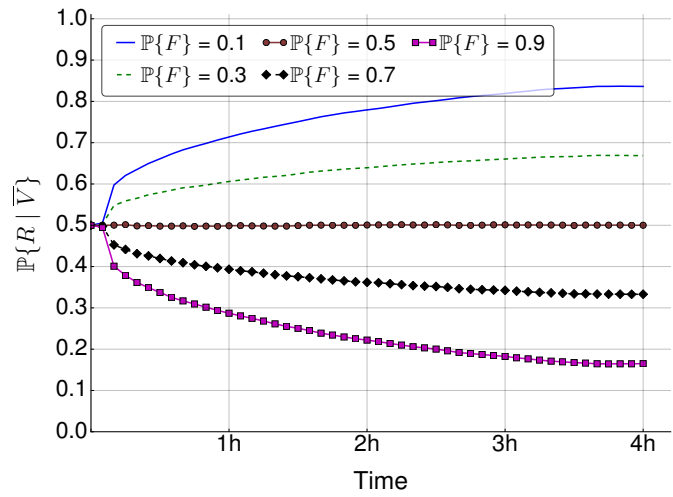

(a)

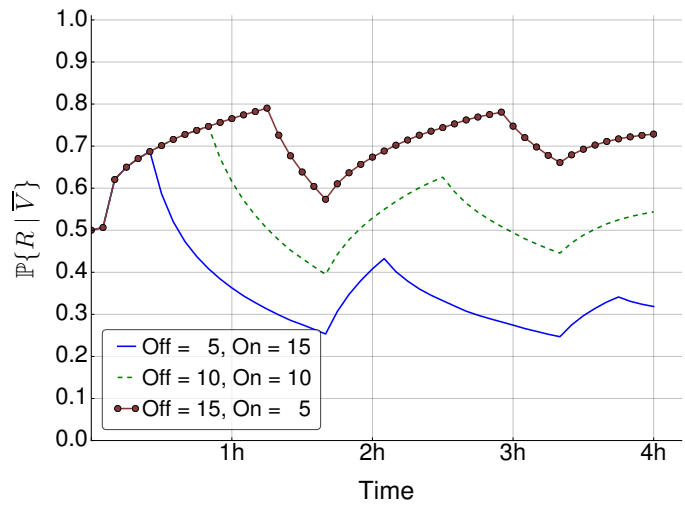

(b)

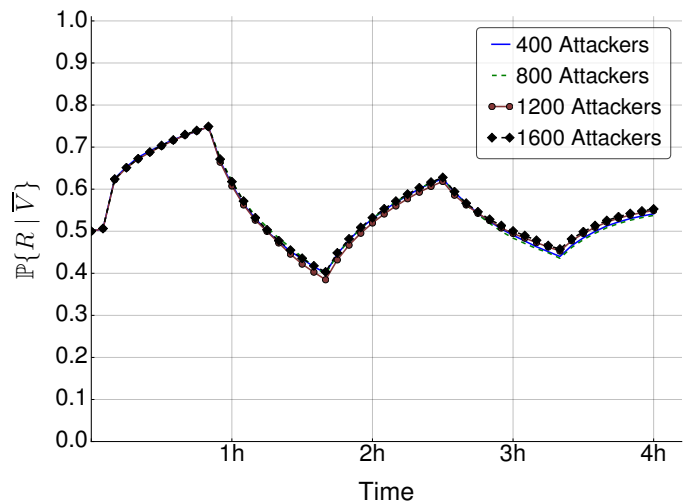

(c)

Fig. 14. FIRST Acceptance probability $\mathbb{P}\{R \mid \overline{V}\}$ in corruption, On-off, and Collusion attacks.

Table 3. Experimental parameters.

| Parameter | Value |
|---|---|
| Experiment length | 240mins |
| Timestep length | 5mins |
| Location | Rome |
| Number of users | 2000 |
| $\mathbb{P}\{F\}$ for non attackers | 0.01 |
| $a_r$ (FIDES) | 0.7 |
| $a_u$ (FIDES) | 0.9 |
| Initial reputation (FIDES) | 0.5 |
| Reputation threshold (FIDES) | 0.75 |
| Initial weights (FIDES) | [1, 0, 0] |
| $\lambda_s$ (Huang) | 0.7 |
| $\lambda_p$ (Huang) | 0.8 |
| $a$ (Huang) | 1 |
| $b$ (Huang) | -2.5 |
| $c$ (Huang) | -0.85 |
| Initial reputation (Huang) | 0.5 |
| Reputation threshold (Huang) | 0.5 |
| Number of MTPs | 400 |
| Number of attackers | 1200 |
| Attackers $\mathbb{P}\{F\}$ | 0.8 |
| On-off steps | (10, 10) |
| Collusion groups | 3 |

Figure 13 shows the results obtained by running the Collusion attack. The experiment has been implemented as follows. We have assumed there are $k$ collusion groups. An attacker belonging to the $k$-th group coordinates with the other attackers belonging to the same group by implementing together an On-off attack. In such attack, during the ON phase the attackers send false reports pertaining to a chosen sector, the same for every user in the $k$-th group. The results conclude that [Huang 2014] and Majority are severely affected by this attack, while FIRST tolerates well this attack by keeping the error rate below 7% by using 400 MTPs, regardless of the number of attackers and collusion groups considered. This is because FIRST uses MTPs to validate data and does not rely on data aggregation. Interestingly enough, [Huang 2014] and Majority perform slightly well when the collusion groups are more. This is explained by considering that when the collusion groups are more, less attackers will belong to the same group, and so it is more likely that a scheme based on aggregation may perform better.

In Figure 14 we report the probability $\mathbb{P}\{R \mid \overline{V}\}$ of FIRST (i.e., the probability that a report will be accepted when not validated) as a function of time, in all the considered attacks. In the Corruption attack, as expected $\mathbb{P}\{R \mid \overline{V}\}$ converges to the $\mathbb{P}\{\overline{F}\}$ probability of the attackers. In the On-off attack,

FIRST reacts by decreasing the $\mathbb{P}\{R \mid \overline{V}\}$ probability and increasing it in the OFF phases. Same behavior is also experimented in the Collusion attack, but in this case, the performance is not affected by the number of attackers as explained above. As described in Section 4.3, $\mathbb{P}\{R \mid V\}$ is equal to $P\{\overline{F}\}$, because when reports are validated the probability of misclassification is zero; on the other hand, when the reports are not validated, we would like that $\mathbb{P}\{R \mid \overline{V}\}$ also tended to $\mathbb{P}\{\overline{F}\}$, and Figure 14 shows that FIRST achieves such goal.

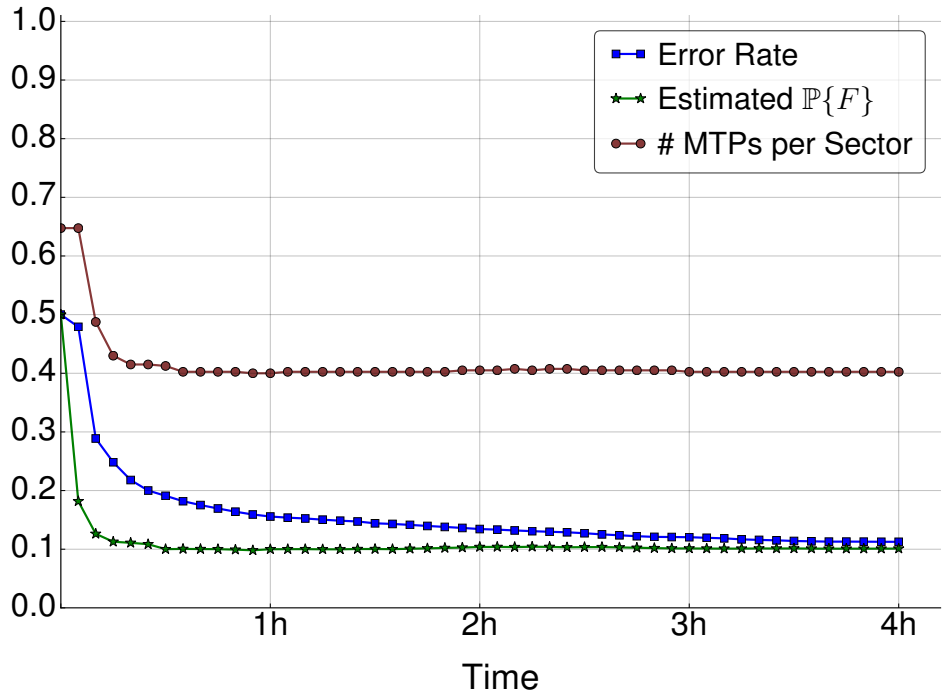

Fig. 15. Error Rate, Estimated $\mathbb{P}\{F\}$ and Number of MTPs per Sector vs time.

Although the MOP was designed to be run before deployment, the MOP may be re-run after deployment by using an estimated $\mathbb{P}\{F\}$ value from the validated data to decrease the number of the MTPs from the worst-case scenario (see the discussions in Section 4.3). To verify this intuition, we have run experiments on the Rome mobility traces [2]. In these experiments, the target classification accuracy of the MOP algorithm is 0.9. First, the MOP is run with $\mathbb{P}\{F\} = 0.5$ to obtain the worst-case value of the number of MTPs, as discussed in R1.*(a)*. As the time goes by, $\mathbb{P}\{F\}$ is estimated every five (5) minutes by using the validated reports as explained in Section 3.2. The new $\mathbb{P}\{F\}$ estimate is then used to run the MOP again and thus decrease the number of MTPs employed. Figure 15 shows the error rate, the estimated $\mathbb{P}\{F\}$ value and the number of MTPs per sector as a function of time. The figure shows that FIRST is able to use the estimated $\mathbb{P}\{F\}$ to decrease the number of MTPs over time (from ∼0.65 to ∼0.4 per sector after seven (7) iterations) without compromising the target classification accuracy (*i.e.*, error rate of 0.1).

## 5.2 Participatory PerCom

In addition to the participatory traffic sensing use-case as described above, we have evaluated the performance of FIRST by implementing an MCS system designed to monitor the attendance of participants at various events during the IEEE PerCom 2015 conference held in St. Louis, Missouri, USA. In such a system, the voluntary participants were asked to regularly submit (i) the conference room they were currently in, and (ii) the (approximate) number of participants in that room. The goal of the experiment was to evaluate the accuracy of FIRST in classifying sensing reports sent by participants in a practical scenario.

*5.2.1* **Experimental setup**. The server-side of the MCS system handling the storage of sensing reports was implemented by using a dedicated virtual machine on *Amazon Web Services*. Figure 16 shows the screenshots of the Android and iOS apps distributed to the participants[1]. The apps provided a simple interface for the participants to report the room they were in (8 choices, from 'A' to 'H'), and the approximate number of people in that room (5 choices, 'Less than 10', 'Between 10 and 20', 'Between 20 and 50', 'Between 50 and 100', and 'More than 100'). In order to recruit participants, we asked the conference and workshop attendees when they picked up registration packages if they were willing to install our app and participate in the experimental study. This way, we were able to recruit 57 participants attending the entire conference and workshops, which is significant considering that we did not incentivize the participants with any kind of reward.

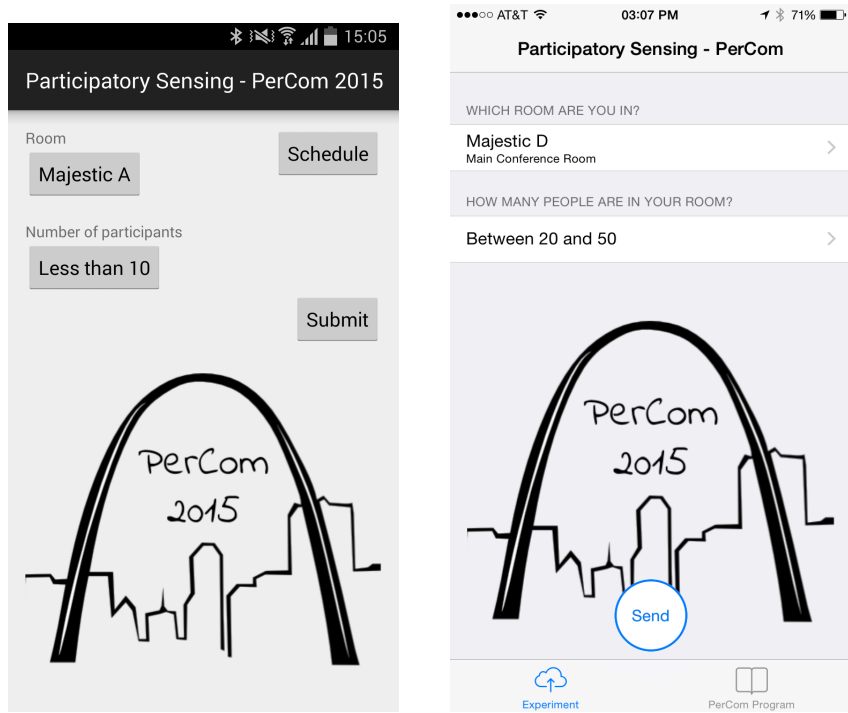

Fig. 16.  Screenshots of the MCS app, Android and iOS.

If yes, they were asked to download and install our app through the `DeployGate`[2] distribution platform. the *IEEE International Conference on Pervasive Computing and Communications* (PerCom), held in St. Louis, Missouri, on March $23^{rd}$-$27^{th}$, 2015.

In order to acquire ground-truth information about the location of participants, we used 20 *Gimbal*™ beacon devices (available at http://www.gimbal.com) which emitted periodically *Bluetooth* packets that were received by the MCS app (deployment illustrated in Figure 17. Whenever a user sent us a report, the location of the nearest beacon was also automatically included in the report by the MCS app. This way, we were able to acquire ground-truth information on user location. To acquire ground-truth information about the number of people in each room, three people voluntarily acted as MTPs and sent every 5 minutes the actual number of people in each conference room. Prior to the conference, to evaluate the impact of the $T$ parameter (i.e., MTP reporting interval), we implemented 5 concurrent, real-time classification processes, each one taking into account different MTP reporting intervals (10, 15, 20, 25, and 40 minutes), aiming at evaluating the impact of the length of the MTP reporting interval.

---

[1]IRB approval of experiments available on file upon request.
[2]Available at http://www.deploygate.com

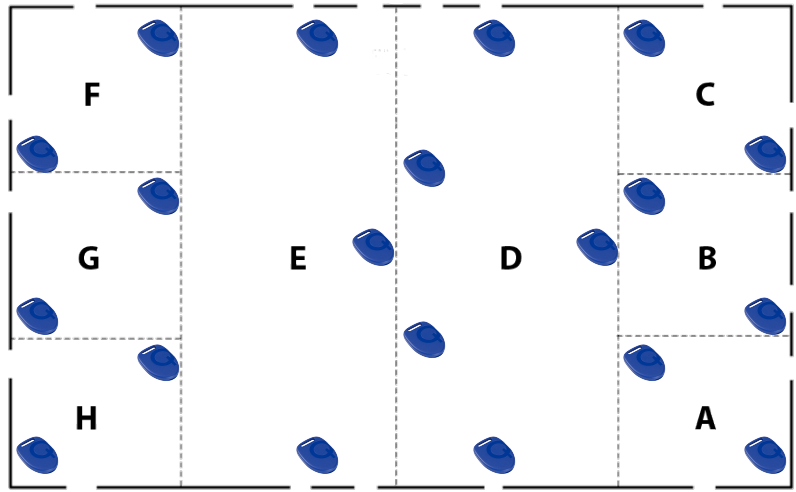

Fig. 17. Position of Bluetooth beacons.

During the experiment, we observed that the number of people attending a particular event was almost constant during a 10-minute time window. Therefore, we used the MTP reports to validate all the reports sent in the following 10-minute time frame. More specifically, we validated a user report as *reliable* if (i) an MTP report $r$ was sent during the 10-minute time frame before the user report was received, *and* (ii) the reported number of people in that room was in the same range as the one sent by the MTP in $r$. If the number of people in the room reported by the user mismatched the information acquired by the MTP, the report was considered *unreliable*. Otherwise, if no MTP report was available during the previous 10-minute time window, we used Equation (11) to decide whether to consider the report as reliable, as explained in Section 4.4. After the experiment, we used the ground-truth information provided by the Bluetooth beacons and the MTPs to calculate the classification accuracy of FIRST.

*5.2.2* **Experimental results**. Figure 18(a) shows the distribution of the percentage of participants that had submitted unreliable reports with a given frequency. For graphical reasons, frequencies in the *x*-axis have been grouped into intervals of length 0.1. Figure 18(a) points out that about 44% of the participants submitted more than 50% of unreliable reports; moreover, over 30% of participants submitted more than 90% of unreliable reports when participating. These results make this experiment ideal to study the performances of FIRST given the number of unreliable reports is significant.

We believe that the reason why some participants reported unreliable information during the Participatory PerCom experiment is that, at the moment of recruiting, we explained that the experiment was part of project aimed at evaluating the robustness of a framework that improves information quality in mobile crowdsensing. Therefore, it is reasonable to assume that some participants were motivated to purposely submit unreliable information, having as objective providing us useful data to stress our FIRST framework and evaluate its resiliency.

Figure 18(b) illustrates the accuracy of the considered approaches as a function of the MTP reporting intervals implemented in the experiments. These results conclude that FIRST outperforms existing approaches as far as classification accuracy is concerned. In particular, FIRST achieves on the average an accuracy of 76.02%, as compared to FIDES, [Huang 2014] and majority vote which achieve 64.45%, 63.99%, and 46.2%, respectively. The results can be explained as follows. When the MTP reporting interval is 10 minutes, both FIDES and FIRST achieve accuracy of 100%, because each report undergoes validation by MTPs. As the reporting interval increases, FIDES performs worse than FIRST due to the challenge in finding a parameter setting which achieves good performance in all scenarios. In contrast, FIRST does *not* require any parameter setting to be implemented, and it is able to achieve high accuracy in all the considered scenarios.

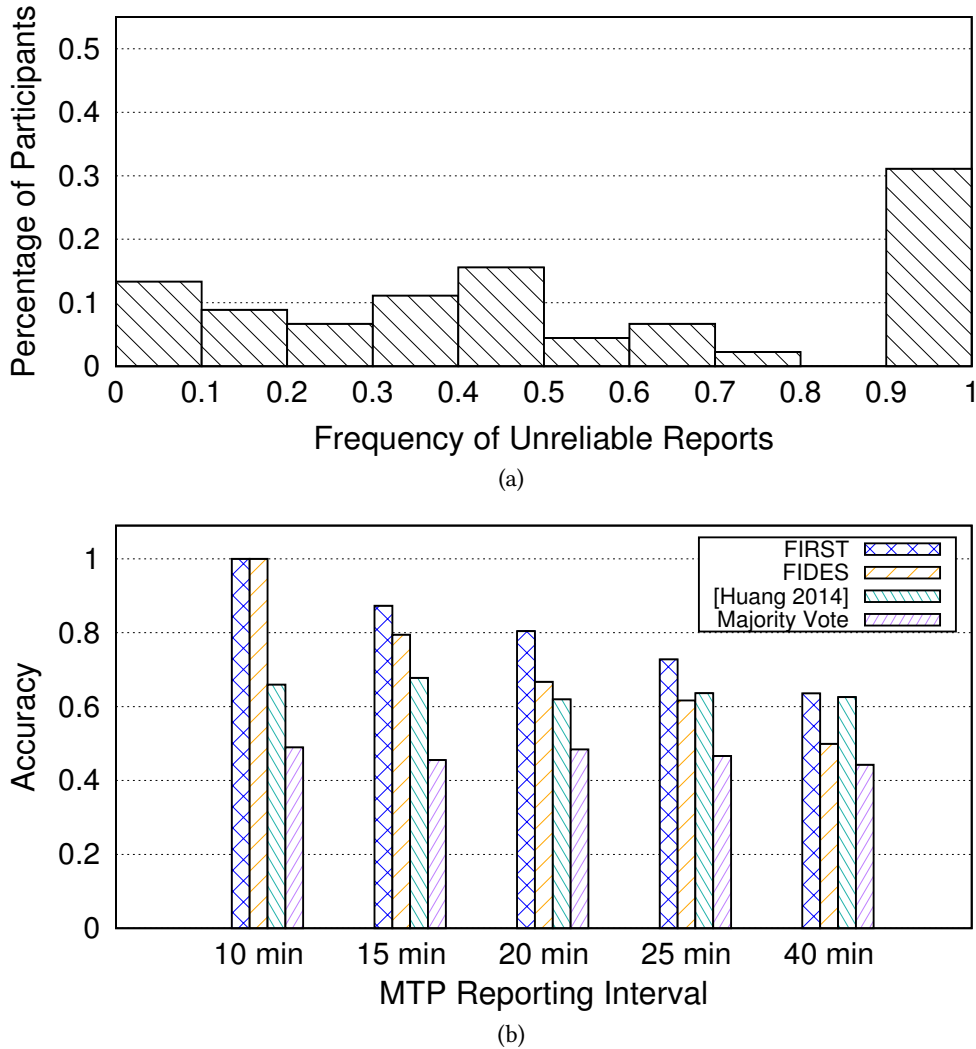

Fig. 18. (a) Frequency of unreliable reports vs. percentage of participants. (b) Comparison of FIRST vs. FIDES, Majority Vote and [Huang 2014].

Note that, as in the traffic sensing experiment, MTP reports were considered as normal users' reports when evaluating Majority and [Huang 2014]. However, given the relatively small number of MTPs involved in this experiment, the accuracy achieved by majority vote and [Huang 2014] in Figure 18(b) does not change significantly when varying the MTP reporting interval. As far as performance is concerned, Figure 18 concludes that such approaches do not obtain accuracy values close to FIRST. This is due to the fact that approaches based on majority vote are not resilient to large number of unreliable reports, which is the case of the Participatory PerCom experiments, as shown in Figure 18(a). It is worth noting that, even in the extreme case of MTP reporting every 40 minutes (which means that only 25% of the reports are verified, on the average) FIRST still outperforms FIDES, Majority and [Huang 2014].

## 6 CONCLUSIONS AND FUTURE WORK

In this paper, we have proposed FIRST, a novel framework that models and optimizes the information reliability in mobile crowdsensing systems. First, we have introduced our system model, the concept of mobile trusted participants (MTPs), and the MTP optimization problem (MOP). Then, we have discussed in details the main components of the FIRST framework, which include a novel likelihood estimation algorithm (LEA) and the MTP optimization algorithm (MOA) that provides optimum solution to the MOP. Furthermore, we have extensively evaluated the framework through real mobility traces in the context of participatory traffic sensing, and by a practical implementation of a system that monitored participants' attendance at the IEEE PerCom 2015 conference. Finally, we have compared FIRST with state-of-the-art literature. Results have shown that FIRST outperforms existing work in increasing information reliability and is able to capture the performance of the system with significant accuracy.

As part of our ongoing work, we are currently working with the U.S. Department of Geological Survey (USGS) to deploy IncentMe along with their *National Map Corps* (TNMCorps) project, which is a U.S. government funded project tasked with mapping our Nation with crowdsourced contributions. Since 2013, the project has expanded to include all 50 states, Puerto Rico, and the US Virgin Islands. TNMCorps volunteers are successfully editing 10 different types of structures, including schools, hospitals, post offices, police stations and other important public buildings. Data from selected structures, including the location and characteristics of manmade facilities are collected by assigning "points" to the structure (i.e., locations of interest). In March 2016, the number of data collection points reached nearly 200,000. Currently, volunteers collect and/or improve a structure's data by adding new points, removing obsolete points, and correcting existing data using web-based mapping tools.

The main challenges of TNMCorps include recruiting, engaging and motivating volunteers. The USGS believes that the addition of mobile crowdsensing to the TNMCorps project would improve dramatically the number of points collected, and help collecting fine-grained and real-time information to an extent which was impossible to achieve in the past. Since August 2015, we have been closely collaborating with the USGS in Rolla, MO. Figure 19 includes a screenshot of the prototype of the app we are developing in collaboration with USGS. The app provides an intuitive interface for participants to update and delete points of interest from the map and retrieve points from the server in real time. The collaboration with the USGS will provide us a platform to test the FIRST framework and other mobile crowdsensing systems to an unprecented extent.

## ACKNOWLEDGEMENT

This material is based upon work supported by the National Science Foundation under grant no. CNS-1545037, CNS-1545050, and DGE-1433659.

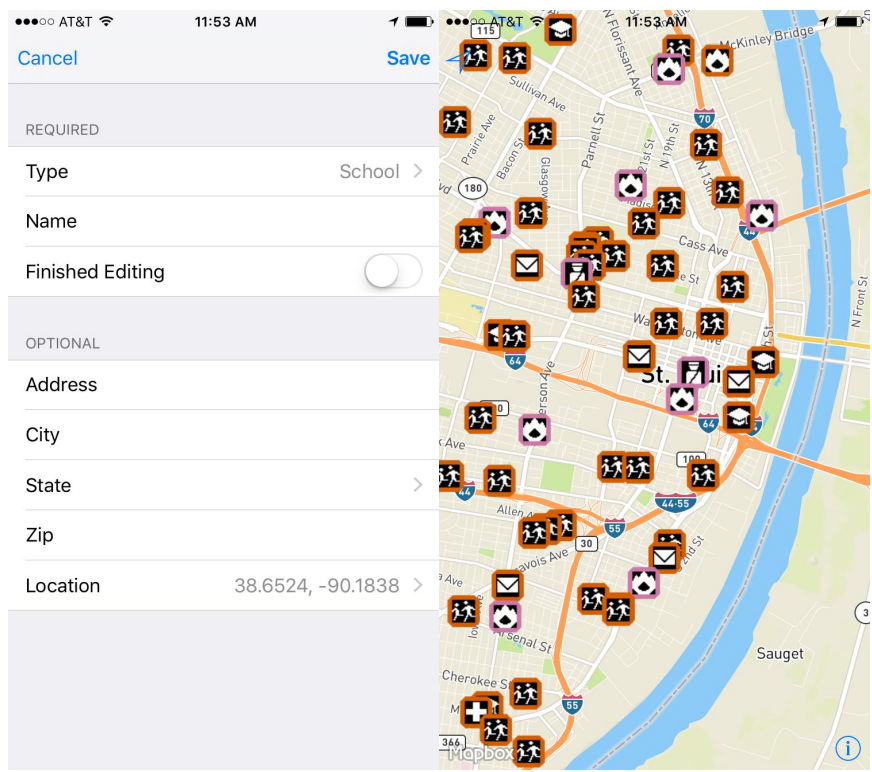

Fig. 19. The National Map Corps App.